\documentclass[aps, prd, amsmath, floats, floatfix, twocolumn,
superscriptaddress, nofootinbib, showpacs]{revtex4}

\usepackage{graphicx}
\newcommand{\beq}{\begin{equation}}
\newcommand{\eeq}{\end{equation}}
\newcommand{\beqn}{\begin{eqnarray}}
\newcommand{\eeqn}{\end{eqnarray}}

\newcommand{\ot}{\overline{t}}
\newcommand{\ox}{\overline{x}}
\newcommand{\oy}{\overline{y}}
\newcommand{\oz}{\overline{z}}

\newcommand{\Caltech}{\affiliation{Theoretical Astrophysics 130-33,
    California Institute of Technology, Pasadena, CA 91125}}
\newcommand{\Cornell}{\affiliation{Center for Radiophysics and Space
    Research, Cornell University, Ithaca, New York, 14853}}

\usepackage{graphicx}% Include figure files
\usepackage{dcolumn}% Align table columns on decimal point
\usepackage{bm}% bold math
\usepackage{epsf}

\begin{document}

\title{Evolving black hole-neutron star binaries in general relativity using
pseudospectral and finite difference methods}

\author{Matthew D. Duez} \Cornell %

\author{Francois Foucart} \Cornell %

\author{Lawrence E. Kidder} \Cornell %

\author{Harald P. Pfeiffer} \Caltech %

\author{Mark A. Scheel} \Caltech %

\author{Saul A. Teukolsky} \Cornell %

%\affiliation{Center for Radiophysics and Space Research, Cornell University,
%  Ithaca, NY 14853}
%\affiliation{Theoretical Astrophysics 130-33, California Institute of Technology,
%  Pasadena, CA 91125}

\begin{abstract}
We present a code for solving the coupled Einstein-hydrodynamics equations
to evolve relativistic, self-gravitating fluids.  The Einstein field equations
are solved in generalized harmonic coordinates on one grid using
pseudospectral methods, while the fluids are evolved on another grid using
shock-capturing finite difference or finite volume techniques.  We show that
the code
accurately evolves equilibrium stars and accretion flows.  Then we simulate an
equal-mass nonspinning black hole-neutron star binary, evolving through the
final four orbits of inspiral, through the merger, to the final stationary
black hole.  The gravitational waveform can be reliably extracted from the
simulation.
\end{abstract}


\maketitle

\section{Introduction}
\label{intro}

Compact object binaries containing neutron stars [i.e. neutron star-neutron
star (NSNS) and black hole-neutron star (BHNS) binaries]
are perhaps as important in modern astrophysics as binary black holes. 
Both BHNS and NSNS binaries should be excellent sources of gravitational
waves for the ground-based interferometric detectors LIGO, GEO, VIRGO,
and TAMA.  It is possible~\cite{2008ApJ...676.1162S} that the detection rate
of these binaries will actually be greater than that of binary black holes.
NSNS and BHNS binaries are also
interesting because they are leading candidates for explaining the production
of short-duration gamma-ray bursts (GRBs)~\cite{1992ApJ...395L..83N},
especially given observations that locate some short GRBs in
elliptical galaxies~\cite{2005Natur.437..851G,2005Natur.437..855V} or rule
out an associated supernova~\cite{2005Natur.437..845F}.  NSNS and BHNS
mergers may also be important for understanding the observed
abundances of the heavy elements that are formed by rapid neutron capture
in the r-process~\cite{1974ApJ...192L.145L}.

NSNS and BHNS binary mergers can be accurately modeled only by numerical
simulations.  In such systems, the spacetime metric and neutron star fluid
are both dynamical.  They are also strongly coupled, and hence must be
evolved simultaneously.  In addition, it is clear from the high compactions
of the binary objects that only simulations in full general relativity will
be adequate.  For NSNS binaries, such simulations are particularly needed to
study the possible collapse of the post-merger remnant.  For BHNS binaries,
the crucial questions are 1) whether or not the neutron star is tidally
disrupted before it plunges into the black hole, and 2) if the neutron star
does disrupt, what fraction of the star's matter is promptly swallowed by
the hole, what fraction is ejected, and what fraction forms an accretion
disk.  To produce a GRB, a substantial accretion disk must remain.  To
contribute to the abundance of r-process elements, matter must be expelled
from the system.  Studies strongly suggest that the non-Newtonian form of
the gravitational potential~\cite{2005ApJ...634.1202R}, gravitational
radiation reaction~\cite{2005ApJ...626L..41M}, and black hole
spin~\cite{Rantsiou:2007ct} all significantly affect the size of the
post-merger accretion disk, underscoring the need for fully relativistic
simulations.

Of the two types of systems, NSNS binaries are better studied.  Newtonian
simulations have been performed using realistic equations of
state together with neutrino radiation effects~\cite{1996A&A...311..532R,
1997A&A...319..122R,1999A&A...341..499R,2002MNRAS.334..481R,2008arXiv0806.4380D}
and magnetic fields~\cite{2006Sci...312..719P}.  Simulations using complicated
equations of state have also been carried out using the conformally flat
approximation to general relativity~\cite{Oechslin:2006uk}.  General relativistic
simulations have been performed using $\Gamma$-law equations of
state~\cite{2000PhRvD..61f4001S,2002PThPh.107..265S,2003PhRvD..68h4020S,
2004PhRvL..92n1101M,2004PhRvD..69f4026M,Baiotti:2008ra} and using more realistic equations
of state~\cite{2005PhRvD..71h4021S,Shibata:2006nm}.  Recently,
general relativistic merger evolutions have been performed that include
the neutron star magnetic field~\cite{2008PhRvL.100s1101A,2008PhRvD..78b4012L}.

Numerical modeling of
BHNS binary mergers has been carried out using Newtonian or
pseudo-Newtonian gravity (in which the black hole is represented by a point
mass) using $\Gamma$-law~\cite{1999MNRAS.308..780L,1999ApJ...526..178L} and
realistic nuclear~\cite{1999ApJ...527L..39J,2004MNRAS.351.1121R,
2005ApJ...634.1202R} equations of state.  BHNS simulations have also been
done in conformal gravity~\cite{2006PhRvD..73b4012F}, although so far only in
the extreme mass ratio limit, in which the black hole is much more massive
than the neutron star.  The first fully relativistic BHNS simulation was
of a head-on collision~\cite{Loffler:2006wa}.   Most recently, two groups
have independently evolved configurations of BHNS binaries in
full general relativity from quasi-circular inspiral through merger.  One group
includes Shibata, Taniguchi, Ury\={u}, and
Yamamoto~\cite{2007CQGra..24..125S,Shibata:2007zm,Yamamoto:2008js}. 
The members of the other group are Etienne, Faber, Liu, Shapiro, Taniguchi,
and Baumgarte~\cite{2008PhRvD..77h4002E}.  The evolutions produced
by these two groups agree qualitatively, with both now finding
very small post-merger disks.  However, both groups have so far restricted
themselves to $\Gamma=2$ polytropic equations of state and to initially
non-spinning black holes. 
The case of spinning black holes has been studied by Rantsiou, Kobayashi, and
Laguna~\cite{Rantsiou:2007ct} using a Kerr background metric and a Newtonian
approximation for the neutron star self-gravity.  They find that black hole spin
has a strong influence on the size of the post-merger accretion disk.  For
both types of binary, but especially BHNS binaries, the parameter space remains poorly
explored.  Also, there is an unmet need for results to be checked by
multiple independent codes.

In each of the above-mentioned calculations in which the metric variables are
truly dynamical fields, these fields were evolved numerically using finite
differencing (FD).  For a stable evolution, such algorithms should converge to
the exact solution as some power of the grid spacing.  Also, shock-capturing
techniques have been developed which allow FD codes to evolve fluids with
discontinuities stably and accurately.  FD codes usually
require very large grids in order to obtain accurate results, although
this problem can be mitigated by using mesh refinement or high-order
differencing.

Einstein's equations can also be evolved using spectral methods.  In this
case, functions are approximated as truncated series
expansions in a set of orthogonal basis functions.  Derivatives of the
approximated functions are then computed exactly.  The pseudospectral (PS)
method is a type of spectral method that uses the values of functions on
a spatial grid of collocation points, rather than directly using the
spectral coefficients.  This has the advantage that pointwise operations
are as straightforward in PS methods as they are in FD methods.  In
{\it multidomain} PS methods, the computational region is divided into
domains, each with its own basis functions and corresponding collocation
points.  For
smooth functions, spectral methods (including PS) converge {\it exponentially}
to the exact solution as the number of basis functions (or, for PS methods,
the number of collocation points) is increased.  This allows PS methods to
get accurate results with much smaller grids than those used by FD codes. 
A PS code for solving the Einstein equations has been developed by the
Cornell-Caltech relativity group~\cite{Kidder:2000yq,Scheel:2002yj,
Lindblom:2005qh,Boyle:2006ne}.  It has been used to simulate the inspiral
of binary black holes for many orbits with very high accuracy for a
fairly low computational cost~\cite{Scheel:2006gg,Pfeiffer:2007yz,Boyle:2007ft}.

There is a difficulty, however, in extending PS methods to evolving non-vacuum
spactimes.  Because of the presence of stellar surfaces
and, in some cases, hydrodynamic shocks, the evolved variables are not
smooth in all derivatives.  In these cases, spectral representations do
not converge exponentially to the exact solution.  Rather, they display Gibbs oscillations
near the discontinuity that converge away only like some power of the
number of collocation points, with the order of convergence given by the order of
the discontinuity.  The oscillations can be controlled using special forms of
filtering or artificial viscosity (e.g.~\cite{t89,d94}) but exponential
convergence is still lost.  In some cases, the problem can be avoided by
placing domain boundaries at discontinuities, e.g. at the stellar
surface.  However, this is not practical for complicated shocks or
when stars become very deformed (e.g. during tidal disruption).

In this paper, we use the ``mixed'' approach developed by Dimmelmeier
{\it et al}~\cite{2005PhRvD..71f4023D} for a conformal gravity
code which has been used quite successfully to study supernova core
collapse~\cite{Dimmelmeier:2006mh,Ott:2006eu,Dimmelmeier:2007ui,
CerdaDuran:2007cr}, and extend it to full general relativity.  In this
method, the spacetime metric is evolved on one grid using PS methods, the fluid
variables are evolved on a second grid using shock-capturing FD or
finite volume techniques, and the two grids communicate by interpolation. 
Below, these two grids are referred to as the ``PS grid'' and the
``fluid grid''.

This approach would seem to utilize the strongest features of each method.  The
spacetime is expected to be smoother than the fluid variables (e.g. at a
stellar surface, the discontinuity in metric components appears at a higher
derivative than in the density), and so PS techniques should work better
on the metric than the fluid.  Also, our code uses many spectral
domains, and we expect discontinuities to appear in only a few domains. 
In the domains without discontinuities, the functions can be represented
spectrally with the same accuracy as in smooth problems.  In the domains
with discontinuities, convergence will be limited to a power law.  Of
course, this error will propagate to the other domains.  However, the
domain decomposition can often be chosen so that the slower converging
domains take up a small fraction of the overall computational region,
and their effect on the overall accuracy is correspondingly small. We
can also use higher resolution in these domains if the error is still
too large.  Unlike the strategy of fitting domain boundaries to
discontinuities, we do not require the exact locations of discontinuities,
but only approximate locations.

Even if the evolution of the fields converges rapidly, spectral accuracy
is still lost because of the evolution of the fluids, which will at best
converge as a fixed power in the grid spacing on the fluid grid.  However,
this is not a problem if the resolution on this grid is high.  And, in fact, our
mixed technique algorithm allows us to achieve high resolution in the
hydrodynamic evolution at a surprisingly low computational cost.  This
is because the fluid grid only needs to cover the region containing the matter. 
This can provide a huge savings for binary inspiral calculations.  For a
BHNS inspiral, the fluid grid can be a box centered on the neutron star with
outer boundaries slightly outside the star.  For a NSNS inspiral, one would
need two such boxes.  For a BHNS system in which the star is shedding mass
onto the black hole, the fluid grid would have to cover a region containing
both binary objects, but even then, it would not have to extend all the
way out into the wave zone, as it would if the metric were being evolved
on the same grid.

In this paper, we describe our code and test its ability to evolve BHNS
binaries.  In Section~\ref{code}, we describe our evolution code.  In
Section~\ref{codetests}, we present tests of this code. 
Next, we apply our code to model the inspiral and merger of a BHNS binary.  We
evolve an equal mass binary with an initially non-spinning black hole and
irrotational neutron star.  Section~\ref{bhnsinspiral} describes the inspiral
calculation, with particular emphasis on the accuracy and rate of convergence
of these simulations.  Section~\ref{bhnsmerger} describes the merger.  Finally,
Section~\ref{conclusions} gives our conclusions and future directions for our work.

\section{Evolution Code}
\label{code}

\subsection{Evolution of the Spacetime}

%\subsubsection{Formalism}

We evolve Einstein's equations using the generalized harmonic
formulation~\cite{1985CMaPh.100..525F,Pretorius:2004jg}.  We use
a first order representation of the system~\cite{Lindblom:2005qh}, in which the
fundamental variables are the spacetime metric $\psi_{ab}$,
its spatial first derivatives $\Phi_{iab}$, and its first
derivatives in the direction normal to the slice $\Pi_{ab}$. 
(Throughout this paper, Latin indices from the first part of the
alphabet $a,b,c,\ldots$ run from 0 to 3, while $i,j,k,\ldots$ run
from 1 to 3.)  From these functions, one can easily extract the
3-metric $g_{ij}$, shift $\beta^i$, lapse $\alpha$, and extrinsic
curvature $K_{ij}$. 
The evolution of the gauge is determined by the gauge source
functions $H_{a} = -\psi^{cd}\Gamma_{acd}$, which
are freely specifiable functions of space and time.  The evolution
equations for $\psi_{ab}$, $\Phi_{iab}$, and
$\Pi_{ab}$ are as given in our earlier paper~\cite{Lindblom:2005qh}, except
we add the matter source term
\begin{equation}
  \partial_t\Pi_{ab} = \cdots - 2\alpha (T_{ab}
  - {1\over 2} \psi_{ab} T^{cd}\psi_{cd}) \ .
\end{equation}

%\subsubsection{Evolution Algorithm}

We evolve $\psi_{ab}$, $\Phi_{iab}$, and $\Pi_{ab}$ using a
multi-domain pseudospectral (PS) code described in earlier
papers~\cite{Kidder:2000yq,Boyle:2006ne,Scheel:2006gg}. 
Each PS domain is either a spherical shell, a cylindrical shell, a cube,
a ``cubed sphere''~\cite{Lehner:2005bz}, a filled sphere, or a filled cylinder. 
Spherical harmonics $Y_{lm}$
are used as angular basis functions on spheres.  Fourier functions
$e^{im\phi}$ are used for the azimuthal direction on cylinders.
Chebyshev polynomials $T_n$ are used as basis functions for each direction
on cubes and cubed spheres, for the radial direction on spherical shells,
and for the radial and $z$-directions on cylindrical shells. 
The basis functions on filled spheres and cylinders must be chosen
specially to have the proper behavior at the origin and axis, respectively. 
For example, for spheres, we must decompose a function $f$ as
$f(r,\theta,\phi) = \sum_{nlm}Q_{nl}(r)Y_{lm}(\theta,\phi)$ (note the
coupled indices), where $Q_{nl}\propto r^l$ near the origin. We use the
functions introduced by  Matsushima and Marcus~\cite{mm95} with $\alpha=1$,
$\beta=1$ for filled cylinders and $\alpha=1$, $\beta=2$ for filled
spheres.  We use boundary conditions~\cite{Rinne:2007ui} designed to prevent the
influx of gravitational radiation and constraint violating metric
perturbations.  Our domain decomposition is chosen to leave an unfilled
(excised) region inside the black hole.  No explicit boundary condition
need be applied at the boundary of the excised region because all
characteristics there flow out of the grid.

%\subsubsection{Interpolation}

The PS grid supplies the fluid grid with the metric fields $\alpha$, $\beta^i$,
$g_{ij}$, $K_{ij}$, $\partial_i\alpha$, $\partial_i\beta^j$, and $\partial_ig^{jk}$. 
These functions must be interpolated from the PS gridpoints to the fluid gridpoints.  This could be
done directly at each fluid gridpoint by summing the values of all the basis
functions at that point with weights given by the known spectral
coefficients. This would, however, be prohibitively expensive, as it
would involve $\sim N_{FD}N_{PS}$ operations, with $N_{FD}$ the number of
destination points (i.e. the number of points on the fluid grid) and $N_{PS}$
the number of points on a domain of the PS grid.  Instead, we used a trick introduced
by Boyd~\cite{b92}.  We first interpolate the metric fields onto
a finer PS grid.  (We usually triple the number of collocation points.) 
This can be done cheaply by switching to spectral space (often an
$N_{PS}\log N_{PS}$ operation procedure), adding more basis functions with
zero coefficients, and switching back to physical space.  Then, one can
get an accurate estimate of the field at fluid gridpoints by doing polynomial
interpolation from the refined PS grid, which takes only $\sim N_{FD}$
operations.  Using this procedure, the CPU time spent on interpolation,
while still significant, is smaller than the time spent on evolving on the
PS and fluid grids.

\subsection{Evolution of the hydrodynamic fields}

%\subsubsection{Formalism}

We model the neutron star matter as a perfect fluid with rest mass
density $\rho$, pressure $P$, specific internal energy $\epsilon$, and 4-velocity
$u^a$, so that the stress tensor is
\begin{equation}
  T_{ab} = \rho h u_a u_b + P \psi_{ab} \ ,
\end{equation}
where $h = 1 + \epsilon + P/\rho$ is the specific enthalpy. 
The evolution of the fluid is determined by the laws of baryon conservation
$(\rho u^a)_{;a}=0$ and energy-momentum conservation $T^{ab}{}_{;a}=0$.  These
give five evolution equations for the variables $D=\alpha\sqrt{g}u^0\rho$,
$\tau=\alpha^2\sqrt{g}T^{00}-D$, and $S_k=\alpha\sqrt{g}T^0{}_k$.  Here $u^0$ is
given by the normalization condition $\psi_{ab}u^au^b=-1$.  The pressure is
given by the equation of state.  Our code is written to evolve general
equations of state of the form
\begin{eqnarray}
  \epsilon &=& \epsilon_{\rm cold}(\rho) + \epsilon_{\rm th} \\
  P &=& P_{\rm cold}(\rho) + (\Gamma_{\rm th}-1)\rho\epsilon_{\rm th}
\end{eqnarray}
(c.f.~\cite{2005PhRvD..71h4021S}).  In this paper, we will mainly use the
simple $\Gamma$-law equation of state:
\begin{eqnarray}
  P_{\rm cold} &=& \kappa\rho^{\Gamma}\\
  \epsilon_{\rm cold} &=& P_{\rm cold}/[\rho(\Gamma-1)]\\
  \Gamma_{\rm th} &=& \Gamma\ ,
\end{eqnarray}
which is equivalent to
\begin{equation}
  P = (\Gamma - 1)\rho\epsilon\ .
\end{equation}

%\subsubsection{Evolution Algorithm}

The hydrodynamic equations have the form
\begin{equation}
\label{flux_eq}
\partial_t u + \partial_i F^i = {\cal S}\ ,
\end{equation}
where $u$ is called a ``conservative variable'' ($D$, $\tau$, or $S_k$,
in our case), while $F^i$ and $\cal S$
are the associated flux and source terms, respectively.  To
solve these equations, one first divides the computational
domain into cells, with one cell associated with each gridpoint. 
Then, from the values of $u$ in the cells, one determines the
fluxes $F^i$ at the interfaces between cells.  From these,
one can compute the net flux into each cell, hence obtaining
the $\partial_i F^i$ term in Eq.~(\ref{flux_eq}).  A specific
hydrodynamics evolution algorithm is determined by a set of
choices: \newline
\newline
1) Discretization

The values of $u$ at gridpoints can be chosen to be the values of
$u$ at the centers of cells ({\it finite difference} discretizaton)
or to be the average values of $u$ inside the cells ({\it finite volume}
discretizaton). The distinction only matters for schemes that are higher
than second-order accurate.  All of our higher order schemes assume
finite differencing except for PPM (see below), for which we have coded
finite difference and finite volume options.  Another discretization
issue is the shape of the grid.  We always use uniformly spaced Cartesian meshes,
i.e., at grid indicies $(i,j,k)$ we have gridpoints (cell centers)
${\bf x}_{i,j,k} = (i\Delta x + x_0, j\Delta y + y_0, k\Delta z + z_0)$.
\newline
\newline
2) Interpolation variables

To compute the fluxes, five independent variables must be interpolated
from cell centers or averages to cell faces.  We interpolate $\rho$ to get face
densities and $u_i$ to get information on face velocities.  We have
experimented with several choices for the fifth variable, which should
carry information about face internal energies.  The pressure $P$ is
a common choice, and we have found that it works adequately.  However,
it may not be ideal for evolving low-temperature stars, the main
application of this paper.  This is because interpolating $P$
does not separate the zero-temperature component of $P$ from the
thermal component.  So, even if $P = P_{\rm cold}(\rho)$ exactly on
all grid values, the interpolated $P$ and $\rho$ will not satisfy
this relationship because of interpolation error, i.e., there is
interpolation ``heating''.  Hence, neutron star simulations carried out
using an isentropic equation of state ($P=\kappa\rho^{\Gamma}$ at all
times) can be very accurate for certain problems
(e.g.~\cite{2002PhRvD..65h4024F,2005PhRvD..71b4035B}). 
The isentropic treatment completely removes spurious heating, but it
is not valid in the presence of shocks.

Another choice would be to evolve $\kappa\equiv P/\rho^{\Gamma}$.  Since
this is a constant for cold stars, it can be interpolated exactly.  The
pressure at faces is then constructed from the interpolated $\kappa$ and
interpolated $\rho$.  We find that interpolating $\kappa$ works well for
cold stars, but behaves badly at shocks.  This can be fixed by switching
the scheme at discontinuities back to interpolating $P$.  Even so, its
usefulness is limited to polytropes.  An even better
choice is to interpolate $P_{\rm th} \equiv P - P_{\rm cold}$, the thermal part
of the pressure.  Then the pressure on the face is the sum of the
interpolated $P_{\rm th}$ and the $P_{\rm cold}$ computed from the
interpolated $\rho$.  With this choice, interpolation error introduces
no heat on the faces for a zero-temperature star.  In addition, it
seems to have no problems evolving shocks, and it can be used with any
equation of state.  By interpolating either $\kappa$ or $P_{\rm th}$,
we find that spurious heating in our stars is often significantly reduced.
\newline
\newline
3) Interpolation method

Interpolation must be done in a special way to accommodate the possibility
of shocks.  (The process is usually called ``reconstruction'' in the
literature.)  Consider a one-dimensional problem, in which we know grid
values $p_i$ of the function $p(x)$.  The reconstruction step involves
computing $p_L = p_{i-1/2-\epsilon}$ and $p_R = p_{i-1/2+\epsilon}$, i.e.
the values of $p$ to the left and right of the grid cell interface. 
(In three dimensions, one must reconstruct at faces in each direction.)  
Many techniques have been developed for interpolating from cell centers
or averages to faces  We have implemented second-order monotonized central (MC)
reconstruction, third-order piecewise parabolic (PPM)~\cite{1984JCoPh..54..174C}
reconstruction, third-order convex essentially nonoscillatory (CENO)~\cite{lo98}
reconstruction, and third-order weighted essentially nonoscillatory
(WENO)~\cite{loc94} reconstruction.  We have found that finite volume PPM
usually gives the best results when evolving equilibrium stars.  We have checked
that this scheme achieves third-order convergence for one-dimensional smooth
flows.  For more general problems, we usually find second-order convergence. 
(See tests below.)
\newline
\newline
4) Flux computation

From $p_R$ and $p_L$, one can compute $u_R$ and $u_L$, the face
values of the conservative variables, and $F_R$ and $F_L$, the
face values of the fluxes.  The ``true'' flux at the interface must
be constructed by some sort of combination of the values on the two sides. 
We use the scheme of Harten, Lax, and van Leer (HLL)~\cite{hll}, in which 
the flux is
\begin{equation}
\label{eq:hll}
  F_{i-1/2} = {c_{\rm min}F_R + c_{\rm max}F_L
  - c_{\rm min}c_{\rm max}(u_R - u_L)\over c_{\rm min} + c_{\rm max}}\ ,
\end{equation}
where $c_{\rm min}$ and $c_{\rm max}$ are the left-going and right-going
sound speeds.  The first two terms in Eq.~(\ref{eq:hll}) provide an
average value of the flux, while the last two contribute to a stabilizing
diffusive term.

The time derivative of the variable $u$ is equal to
\hbox{${\cal S} - {\rm div}F$} (c.f. Eq.~(\ref{flux_eq})).  For a finite volume method,
the flux divergence is just
\begin{equation}
\label{fluxdiv}
  {\rm div}F_i = {F_{i+1/2}-F_{i-1/2}\over \Delta x}
\end{equation}
plus the equivalent terms for the other two dimensions.  For a finite
difference method, other terms must be added to achieve third-order
accuracy (see e.g.~\cite{2002A&A...390.1177D} for details).

The vacuum region outside the star or stars requires special treatment. 
We fill the vacuum with a very low density ($~10^{-7}$ of the maximum
value of $\rho$) atmosphere, as is usually done for these types of
problems.  At each step, we apply a floor on $\rho$.  We also apply
both a floor and a ceiling on the temperature in the atmosphere,
requiring $0 < T < 8(P_{\rm cold}/\rho)$.  Finally, we limit the velocity in
the atmosphere to some fraction (e.g. 0.8) of the speed of light. 
The pressure and velocity cutoffs are only applied in regions where
the density is below a chosen threshold (usually $\sim 10^{-3}$ of the
maximum value of $\rho$).  Therefore, they should have little effect on evolutions.

%\subsubsection{Interpolation}
The PS grid requires $T_{ab}$ at each collocation point, so we must
interpolate from the fluid grid to the PS grid.  We have written
an arbitrary-order polynomial interpolator and a third-order WENO
interpolator.  One would expect the latter to give more sensible
results near strong shocks, but for the applications reported here,
both work sufficiently well.

When evolving black hole systems, we excise the gridpoints in a neighborhood
of the singularity and evolve the region near the excision zone using one-sided
differencing as described in~\cite{Hawke:2005zw}.  Of course, the excised region of the
fluid grid must match that of the PS grid.  Also, when interpolating from the
fluid to the PS grid, we choose interpolation stencils which avoid the excised
region.

When symmetries are present in a problem, they can sometimes be used
to reduce the size of the fluid grid.  We have coded options
to impose reflection symmetries on the $x=0$, $y=0$, or $z=0$ planes, or
any combination of them.  In these cases, the fluid grid covers only part of the system,
with fluid quantities in other regions being filled for the PS grid using
the appropriate symmetries.  We have also coded a version of the
hydrodynamics code that assumes axisymmetry and uses a two-dimensional
fluid grid.

Both the PS and the hydrodynamics modules of our code use multiple domains. 
This allows the code to be run in parallel by assigning different domains
to different processors.  We place ghost zones at processor interfaces in
the fluid grid, so that numerical derivatives can be correctly taken at
all true grid points.

\subsection{Evolution of the coordinates}
\label{dual_frames}

Accuracy in a multidomain PS code strongly depends on having
a domain decomposition which matches the ``shape'' of the fields
being evolved.  Also, the excised region of a black
hole must remain inside the horizon.  Thus, it is desirable for
the grid to move with the black holes and neutron stars.  We do
this using the ``dual coordinate frames'' system developed for binary black
hole evolutions~\cite{Scheel:2006gg}.  The coordinate frame $x^{\bar{\imath}}$
is set to be an asymptotically flat, inertial frame.  All tensor components
are evaluated with respect to this frame.  The PS and fluid gridpoints are
fixed in the computational frame $x^i$.  By means of a mapping between the
frames, the computational coordinates approximately co-move with the system. 
For example, one can track a binary using a
simple combination of rotation and radial scaling:
\begin{eqnarray}
\label{RotAndScale}
\begin{split}
%\begin{array}{l}
\ot &= \ot \\
\ox &= x_c + a[(x - x_c)\cos(\Phi) - (y - y_c)\sin(\Phi)] \\
\oy &= y_c + a[(x - x_c)\sin(\Phi) + (y - y_c)\cos(\Phi)] \\
\oz &= az   \ ,
%t &= \ot \\
%x &= x_c + a[(\ox - x_c)\cos(\Phi) + (\oy - y_c)\sin(\Phi)] \\
%y &= y_c + a[(x_c - \ox)\sin(\Phi) + (\oy - y_c)\cos(\Phi)] \\
%z &= a\oz   \ ,
%\end{array}
\end{split}
\end{eqnarray}
where $(x_c,y_c,0)$ is the point at which $x^i = x^{\bar{\imath}}$, which we
choose to be fixed for all time, while $\Phi$ and $a$ are functions
of time.  One evolves $\Phi$ to correct for the binary's orbital rotation,
and $a$ to correct for any change in the orbital separation (i.e. inspiral
or eccentricity).  This is done by introducing a feedback
mechanism into the evolution of $\Phi$ and $a$ (see~\cite{Scheel:2006gg}
for details).

The dual frame mappings actually used in our simulations are
very similar to Eq.~(\ref{RotAndScale}) with one notable difference. 
For large outer radius $R_0$, small changes in $a$ lead to large absolute changes
in $aR_0$, causing an instability.  Therefore, we replace the linear scaling
${\overline r} = ar$ used in Eq.~(\ref{RotAndScale}) with
${\overline r} = ar + (1 - a)r^3/R_0^2$, so that we recover the linear scaling
when $r\ll R_0$, but the outer boundary is fixed even when $a$ changes.  Such
a mapping has been successfully used for long-inspiral binary black hole
simulations~\cite{Boyle:2007ft}. 
Hereafter, we will refer to this coordinate map, a rotation combined with
a nonlinear scaling, as $\mathcal{RS}$.

Since the equations we integrate are written in the inertial frame,
they involve derivatives with respect to $x^{\bar{\imath}}$.  These
derivatives are obtained by first taking derivatives with respect
to grid coordinates $x^i$ and then transforming them using a
Jacobian matrix.  So, for example the equation
\begin{equation}
\partial_{\overline{t}}u^{\overline{\alpha}}
    + A^{\overline{k\alpha}}{}_{\overline{\beta}}
    \partial_{\overline{k}}u^{\overline{b}} = 0
\end{equation}
becomes
\begin{equation}
\partial_tu^{\overline{\alpha}}+\left[
%\beta^i_A
{\partial x^i\over \partial\overline{t}}
	\delta^{\overline{\alpha}}{}_{\overline{\beta}}
      + {\partial x^i\over\partial x^{\overline{k}}}
      A^{\overline{k\alpha}}{}_{\overline{\beta}}\right]
    \partial_iu^{\overline{\beta}} = 0\ .
\end{equation}
%where $\beta^i_A = {\partial x^i\over \partial\overline{t}}$
Both the metric and the fluid can be evolved in
this way.  Thus Eq.~(\ref{flux_eq}) becomes
\begin{eqnarray}
\label{advfluxeq}
\begin{split}
  \partial_tu +  {\partial x^i\over\partial x^{\overline{k}}}
  & \left[\partial_i\left(
%    \beta_A^{\overline{k}}
    {\partial x^i\over \partial\overline{t}}{\partial x^{\overline{k}}\over\partial x^i}
    u + F^{\overline{k}}\right)\right.
\\
   & - u \left.
%    \partial_i\beta_A^{\overline{k}}
    \partial_i\left({\partial x^i\over \partial\overline{t}}{\partial x^{\overline{k}}\over\partial x^i}\right)
    \right]
    = {\cal S}\ .
\end{split}
\end{eqnarray}
Note that we have moved the coordinate advection term
${\partial x^{\overline{k}}\over\partial x^i}{\partial x^i\over \partial\overline{t}}u$ into the derivative containing
the flux so it can mostly cancel the velocity advection
term in $F^{\overline{k}}$.

There are some subtleties involved in applying
dual frames to our shock-capturing techniques; a
straightforward evolution of Eq.~(\ref{advfluxeq}) will often
be unstable.  However, we have implemented two techniques
for evolving fluids in moving coordinates that have proven
to be robustly stable.  We believe that the advantages of moving
coordinates are so great that it is useful to report both methods.

The conceptually simpler technique is to transform the metric and
the field variables into moving coordinates, and then evolve as usual
in this coordinate system. One must remember that $D$, $\tau$, and $S_k$
are densities, so transforming them to the moving frame involves multiplying
by $\det(\partial x^i / \partial x^{\overline{k}})$.  Since the spectral
code evolves tensors using inertial frame components, coordinate
transformations must be performed when the two grids communicate.

It is also possible to evolve the hydrodynamic equations with inertial
frame components using Eq.~(\ref{advfluxeq}).  However, one must make a
few adjustments to the shock-capturing code.  1) The
$c_{\rm min}$ and $c_{\rm max}$ in Eq.~(\ref{eq:hll}) should
be computed in the moving frame.  2)  The fluxes in
Eq.~(\ref{eq:hll}) must be broken up into two pieces.  For
example, for the $x$-interface fluxes
\begin{eqnarray}
\label{f1}
F^{\overline{k}(1)}_{i+1/2} &=& 
(c_{\rm min}+c_{\rm max})^{-1}(c_{\rm min}F^{\overline{k}}_R
+ c_{\rm max}F^{\overline{k}}_L)\\
F^{x(2)}_{i+1/2} &=& (c_{\rm min}+c_{\rm max})^{-1}
c_{\rm min}c_{\rm max}(u_R - u_L)\ . 
\end{eqnarray}
The $\overline{k}$ index may seem unfamiliar; it is not usually needed because
in evolutions without dual frames, one only needs the $x$-derivative of $F^x$,
the $y$-derivative of $F^y$, and the $z$-derivative of $F^z$.  However,
the Jacobian in Eq.~(\ref{advfluxeq}) mixes derivatives, so the other derivatives
are now needed.  To the two flux pieces correspond two pieces of the flux derivatives
\begin{eqnarray}
(\partial_x F^{\overline{k}})_i^{(1)} &=& \Delta x^{-1}(F^{\overline{k}(1)}_{i+1/2} - F^{\overline{k}(1)}_{i-1/2})\\
(\partial_x F^x)_i^{(2)} &=& \Delta x^{-1}(F^{x(2)}_{i+1/2} - F^{x(2)}_{i-1/2})\ ,
\end{eqnarray}
and similarly for $y$ and $z$.  Next, the Jacobian in Eq.~(\ref{advfluxeq}) is
applied to $(\partial_i F^{\overline{k}})^{(1)}$ to give
$(\partial_{\overline{i}} F^{\overline{k}})^{(1)}$.  The Jacobian is not applied to
$(\partial_i F)^{i(2)}$; this term is added directly to the time derivative of $u$. 
If the Jacobian is applied to $(\partial_i F)^{i(2)}$, it will not behave correctly as
a diffusion term, and the code will be unstable.

\section{Code Tests}
\label{codetests}

\subsection{Shocks}

\begin{figure}
\includegraphics[width=8cm]{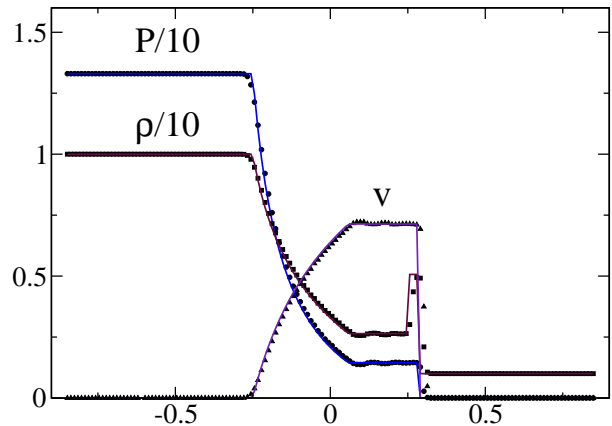}
\caption{A snapshot at $t=0.35$ of the density, pressure, and
  3-velocity at each gridpoint on a line perpendicular to the shock
  interface.  The lines show the exact solution.  Circles, squares
  and triangles show the numerical pressure, density, and velocity,
  respectively.
}
\label{fig:shock}
\end{figure}

As a first test of the hydrodynamics code, we perform a standard
Riemann shock tube problem.  We evolve a $\Gamma=5/3$ fluid with an
interface separating regions of different $\rho$ and $P$.  The initial
state is $(\rho,P,v)=(10,13.3,0)$ on one side of the interface and 
$(\rho,P,v)=(1,0,0)$ on the other.  The interface is chosen to lie
along the $xy$ diagonal, so the problem is numerically two-dimensional. 
We evolve using the PPM reconstructor on a grid of $150^2$ points and
grid spacing $\Delta x = 0.013$.  The results at
$t=0.35$ are shown in Fig.~\ref{fig:shock}. We see that our code captures
all of the features of the exact solution.

\subsection{Spherical Accretion}

Next, we check that our code can accurately model accretion onto a
nonrotating black hole.  Note that accretion test problems provide very
sensitive checks of a code's treatment of the effects of curved spacetime
on the fluid flow (the non-Minkowski pieces of the fluid equations), and also
checks the code's ability to advect matter through an excision boundary
without losing stability or accuracy.

\begin{figure}
\includegraphics[width=7.3cm]{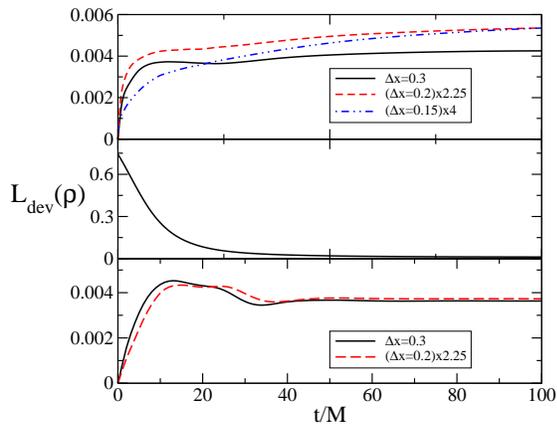}
\caption{ The deviation of the density $\rho$ from its exact
  equilibrium $\rho_{\rm ex}$.  $L_{\rm dev}(f)$ is a relative L2 norm:
  $L_{\rm dev}(f) = L2(f-f_{\rm ex})/L2(f_{\rm ex})$, where the norms sum only
  over points outside the horizon.  The top panel is a relativistic
  Bondi flow, the middle panel is a perturbed Bondi flow, and the bottom
  panel is a moving black hole flow.  The deviations in the top and
  bottom panels are scaled for second-order convergence.
}
\label{fig:bondi}
\end{figure}

We perform three types of accretion tests.  For the first test, we check that
the code can maintain an equilibrium radial accretion flow.  As an initial state,
we use the well-known exact solution of Michel~\cite{1972Ap&SS..15..153M}.  We use Kerr-Schild
coordinates and excise all gridpoints inside a radius of $r=1.6M$, where $M$ is
the mass of the black hole, so the event horizon is at $r=2M$.  We choose a $\Gamma=5/3$
fluid and a flow with the sonic point at $r_s=8M$.  As an outer boundary condition,
we hold the fluid variables at $r>12M$ to the exact values.  The metric fields are held
fixed---otherwise, the black hole would grow, and the problem would not be
stationary.  Derivatives of the metric are computed analytically. 
The fluid is evolved to $t=100M$ on three grids,
with grid spacings of $\Delta x = 0.3M$, $\Delta x = 0.2M$, and $\Delta x = 0.15M$. 
We assume that octant symmetry is maintained, and so evolve only one octant.  The
results are shown in the top panel of Fig~\ref{fig:bondi}.  The solution remains accurate
everywhere, including near the sonic point.  We find slightly worse than second-order
convergence.  This is because of the nonsmooth flow at the sonic point.  We
find that in most of the smooth regions, the flow converges to the exact solution at
second order.

For the second test problem, we perturb the initial state by multiplying the density by
1-$e^{-(r/(6M))^3}$.  We then evolve with the same outer boundary condition as before. 
The flow should relax to the known equilibrium solution, and one can see from the middle
panel of Fig~\ref{fig:bondi} that it does.

Finally, we simulate a case in which the black hole is moving relative to the fluid at
large distance.  We choose the fluid to have the equation of state $P=\rho$.  For this equation
of state, the equilibrium has been computed analytically by Petrich, Shapiro, and
Teukolsky~\cite{1988PhRvL..60.1781P}.  We take the relative velocity between the hole and the
distant fluid
to be $0.6c$ in the $z-$direction, and we evolve in the black hole's frame.  The fluid is evolved
at two resolutions, corresponding to $\Delta x = 0.3M$ and $\Delta x = 0.2M$.  Because of the
front-back asymmetry, we must evolve two octants.  The result is shown in the bottom panel
of Fig~\ref{fig:bondi}.  The convergence is second-order.  We note that for this equation of
state, the speed of sound is equal to the speed of light, so there is no sonic point, and
the solution is smooth everywhere.

\subsection{Equilibrium Relativistic Stars}

\newcommand{\stara}{A}
\newcommand{\starb}{B}
\newcommand{\starc}{C}
\newcommand{\stard}{D}
\newcommand{\ee}[2]{\ensuremath{#1\times 10^{#2}}}

In the previous tests, the matter flow did not affect the spacetime
metric.  For our next test, we evolve equilibrium relativistic stars. 
Self-gravity is obviously of fundamental importance for such objects,
so the fluid and spacetime should now be evolved together, and we
expect significant feedback in both directions.  For this test, we
evolve isolated stars with polytropic equation of state
$P=\kappa_0\rho^{1+1/n}$, where we choose $n=1$.  We choose units
such that $G = c = \kappa_0 = 1$.  In these units, the maximum mass
for a nonrotating $n=1$ polytrope is $M_{\rm max TOV} = 0.164$.  This
is the mass of a polytrope that has the critical central density
$\rho_{\rm crit} = 0.32$.  Equilibrium configurations for rotating
stars were computed using the code of Cook, Shapiro, and
Teukolsky~\cite{1992ApJ...398..203C}.

\begin{table}
%\caption{Equilibrium Polytropes ($n = 1$, $\kappa = 1$).}
\begin{tabular}{cccccccc}
\hline \hline  
   Star
 & $M/M_{\rm max TOV}$
 & $R_{\rm polar}/R_{\rm equat}$
 & $T/|W|$
 & $\Omega_{\rm equat}/\Omega_{\rm axis}$
 \\ \hline  
\stara\ & 0.96 & 1.00 & 0.00 & NA  \\
\starb\ & 0.99 & 1.00 & 0.00 & NA  \\
\starc\ & 0.96 & 0.75 & 0.06 & 1.0 \\
\stard\ & 1.70 & 0.30 & 0.25 & 0.3 \\
\hline \hline  
\label{table:stars}
\end{tabular}
\caption{
  For each of the equilibrium polytropic stars used to test
  the code, the ratio of the ADM mass $M$ to the TOV maximum
  mass $M_{\rm max TOV}$, the ratio of the polar
  ($R_{\rm polar}$) to the equatorial ($R_{\rm equat}$)
  coordinate radius, the ratio of
  the rotational kinetic energy $T$ to the gravitational
  potential energy $|W|$, and the ratio of the angular
  velocity on the equator $\Omega_{\rm equat}$ to the
  angular velocity on the rotation axis $\Omega_{\rm axis}$.
}

\end{table}

We study four stars, whose properties are summarized in Table~I. 
Stars~\stara\ and~\starb\ are nonrotating.  Star~\stara\ has central
density $(2/3)\rho_{\rm crit}$, and star~\starb\ has central density
$(4/3)\rho_{\rm crit}$.  From the turning point theorem~\cite{Friedman:1988er}, we
infer that star~\stara\ is stable and star~\starb\ is unstable. 
These stars have octant symmetry and are evolved using fluid grids
that cover one octant.  Stars~\starc\ and~\stard\ are rotating, and we
evolve them using fluid grids that cover the upper half-plane.  Star~\starc\ has
the same rest mass as star~\stara, but it rotates rigidly at an angular
speed of 70\% its mass shedding limit.  From the turning point theorem,
we know this star is secularly stable, and we expect it to be dynamically
stable as well.  Star~\stard\ is a differentially
rotating, hypermassive star similar to one studied by other
groups~\cite{Baumgarte:1999cq,Duez:2005sf}.  It is known from the
simulations of these groups to be dynamically stable.

\begin{figure}
\includegraphics[width=7.3cm]{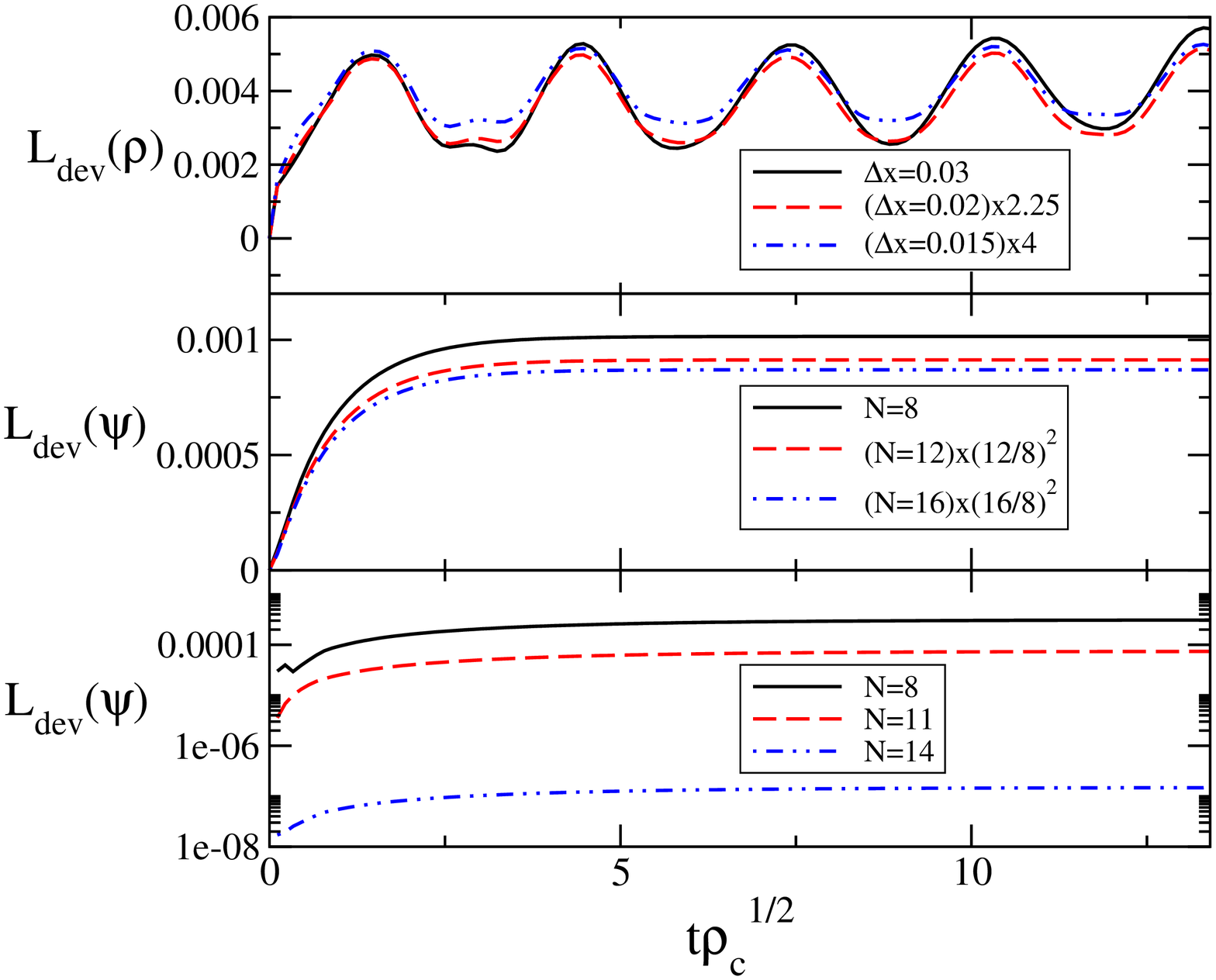}
\caption{
 The deviations from known equilibrium values of the density $\rho$
 and 4-metric $\psi_{\alpha\beta}$ for various grids.  As in
 Fig~\ref{fig:bondi}, $L_{\rm dev}(f) = L2(f-f_{\rm ex})/L2(f_{\rm ex})$. 
 In the top panel, the metric is fixed and the fluid is evolved.  In the
 other panels, the fluid is fixed and the metric is evolved.  In the middle
 panel, the computational domain contains the stellar surface.  In the bottom
 panel, the domain is entirely inside the star.
}
\label{fig:stars_convg}
\end{figure}

First, we investigate the convergence of the fluid and PS codes, and in
particular the effect of the discontinuity in the gradient
of $\rho$ on the convergence of the PS code.  For this, it is sufficient
to look at the spherically symmetric star~\stara\ configuration.  This
star has a nonzero density at any point whose coordinate distance $r$ from
the star's center is less than the stellar radius $R$ in these coordinates. 

We carry out three sets of convergence tests, with results shown in
Fig~\ref{fig:stars_convg}.  First, we hold the metric fixed and evolve the
fluid using three resolutions of the fluid grid.  As expected, we see
second-order convergence.  Because we plot the L2 norm of the deviation
of $\rho$ from its initial value, the frequency of oscillations seen in
the plot is twice the fundamental radial mode frequency.

Next, we hold the hydrodynamic variables fixed
and evolve the metric using the PS code.  We fix the generalized harmonic
gauge source functions $H_a$ at their initial values, which are chosen to
make the spacetime stationary in our coordinates.  To study the effects of the
surface, we do two sets of runs.  First, we evolve the metric on a domain
that contains the stellar surface in the domain interior.  The domain chosen is a
spherical shell with inner radius at $0.7R$ and outer radius at $1.3R$. 
The middle panel of Fig~\ref{fig:stars_convg} shows the results for PS grids
with 8, 12, and 16 radial collocation points.  The nonsmoothness of
the metric at the surface reduces the convergence to second order. 

Finally, we evolve the metric in a PS domain in the interior of the star. 
The domain chosen is a cube at the center of the star with sides of length
$0.86R$.  We evolve with 8, 11, and 14 collocation points in each direction. 
The resulting error is plotted in the bottom panel of Fig~\ref{fig:stars_convg}. 
The error should decrease exponentially as the number of points is increased,
and we do see extremely rapid convergence.

\begin{figure}
\includegraphics[width=7.3cm]{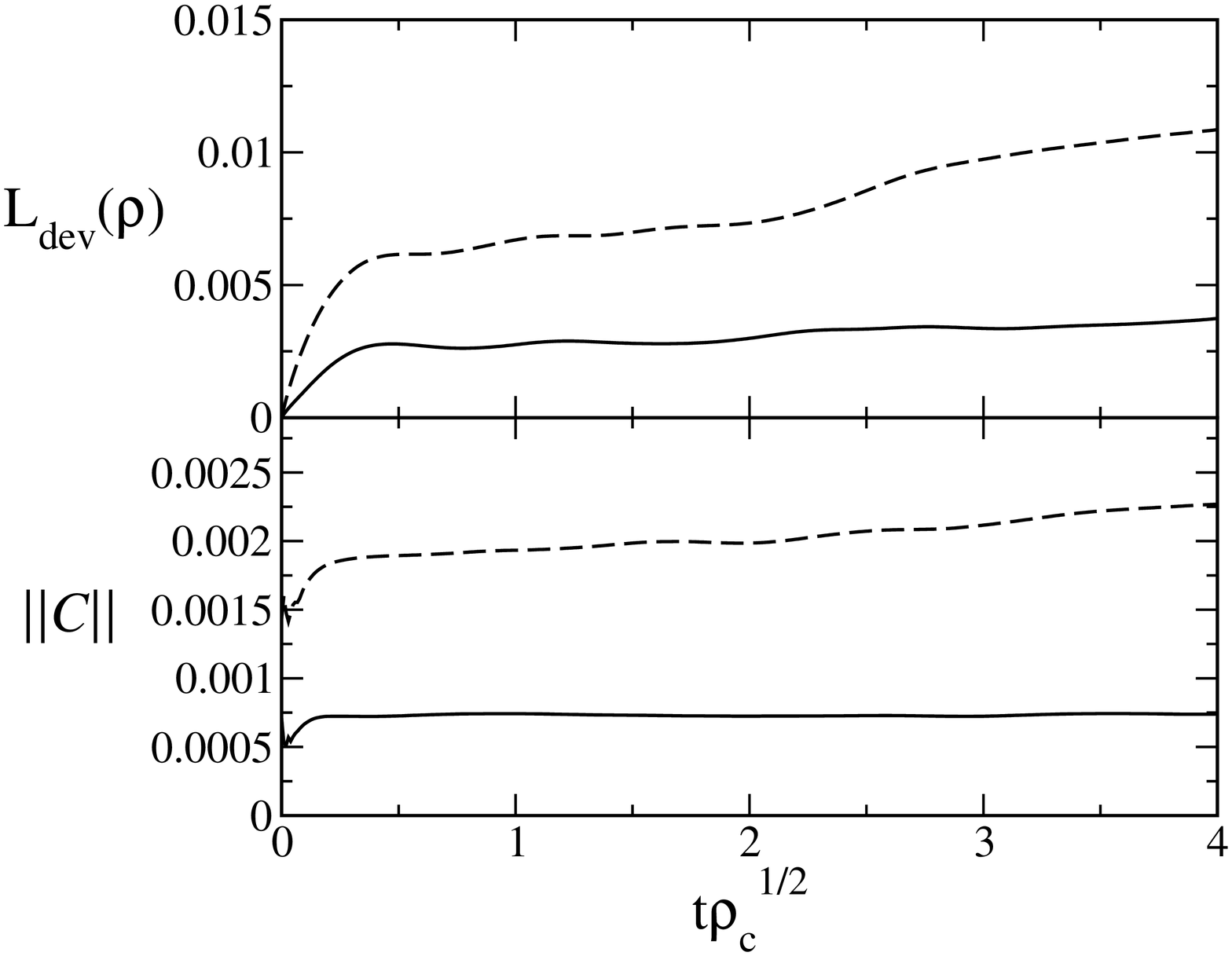}
\caption{ A neutron star with a realistic equation of state with both
  matter and metric evolved.  Two runs are shown, corresponding to lower
  resolution PS and fluid grids and higher resolution grids.  We plot the
  normalized L2 norm of the change in density and the normalized L2 norm
  of the violation in the generalized harmonic constraints.}
\label{fig:sly}
\end{figure}

To test the full code, we next evolve a star in time, allowing both the hydrodynamic fields
and the metric to change and influence each other.  First, we check that
the code still converges.  To demonstrate the generality of the
hydrodynamics code, we pick a more complicated equation of state for the
star used in this test.  (Of course, we have also checked that this test works
with the polytrope star~\stara.) We choose the SLy neutron star equation of
state~\cite{2001A&A...380..151D} to describe $P_{\rm cold}$,
using the fitting function approximation presented in~\cite{2005PhRvD..71h4021S},
and we choose $\Gamma_{\rm th} = 2$.  The star has a central density
of $10^{15}$ g cm${}^{-3}$ and a mass of 1.4$M{}_{\odot}$.  In Fig.~\ref{fig:sly},
we show results from evolving this star at two resolutions.  In the low
resolution run, the fluid grid has $16^3$ gridpoints covering one octant, while the PS
grid is an $11^3$ cube surrounded by 7 spherical shells, each with
6 radial collocation points and a maximum angular $l$ of 8.  In the
high resolution run, the fluid grid has $24^3$ points, while for the PS grid
we add one basis function in each direction in each domain.  Fig.~\ref{fig:sly}
shows the deviation of the density from its initial profile and the
violation in the generalized harmonic constraint equations, the latter
measured by the normalized constraint energy $\|\mathcal{C}\|$
defined by Eq.~(71) in~\cite{Lindblom:2005qh}).  We see that both measures
of error decrease significantly as resolution is increased.  (In this
case, we have no expected convergence rate for comparison.  However,
we note that in Fig.~\ref{fig:sly}, $L_{\rm dev}(\rho)$ decreases as if
it were converging to second-order in fluid grid spacing.)  

\begin{figure}
\includegraphics[width=7.3cm]{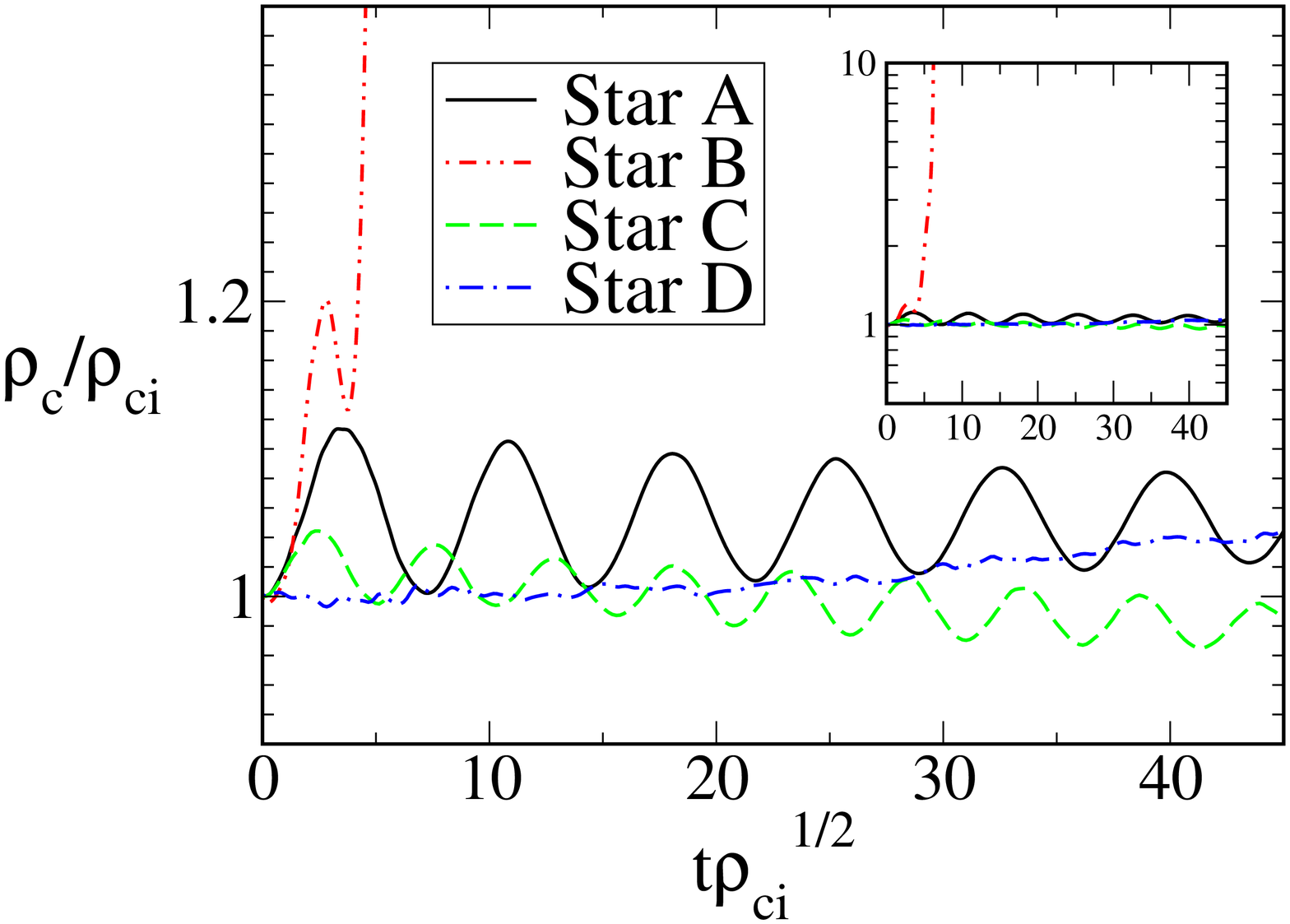}
\caption{ The central density (relative to its initial value) for
  simulations of the four isolated stars.  At the initial time,
  each star is subjected to a perturbation in the form of a one
  percent pressure depletion throughout the star.  The inset shows
  the same thing but with a larger density range, so that later stages
  of the collapse of star~\starb\ can be seen.
}
\label{fig:star}
\end{figure}

As a final test, we evolve all of the polytropic stars
(\stara\, \starb\, \starc\, and D) for many dynamical times
to check that our code can distinguish stable from unstable stars
and can evolve stable stars accurately for long times.  We evolve stars
\stara\ and \starb\ using octant symmetry to restrict the fluid evolution
to a single octant.  For these stars, we divide the computational domain
into 8 fluid domains with $18^3$ gridpoints each and 8 PS domains with roughly
$10^3$ collocation points on an average domain.  For stars \starc\ and
\stard, we utilize equatorial symmetry to restrict the fluid evolution to
the upper half-plane.  We use 32 fluid domains of $19^3$ gridpoints and
32 PS domains of roughly $10^3$ points each.  For all four stars, the
PS grid consists of a cube at the star's center, a spherical shell well
outside the star at the outer boundary, and the region in between covered
by several layers of cubed spheres, one of which contains the stellar
surface.  The generalized harmonic
gauge source functions $H_a$ are again chosen to be fixed in time at their
initial values.  In order to test the
stability of each star, we perturb the initial data by reducing the
pressure everywhere by one percent.  The induced violation of the
constraint equations is small, so we do not adjust the metric to re-solve
the constraints.

In Figure~\ref{fig:star}, we plot
the central density as a function of time for each star.  As anticipated,
stars \stara, \starc, and \stard\ are stable, while star \starb\ collapses. 
(We do not attempt to follow the collapse of star \starb\ to late times
because we expect our choice of $H_a$ to be poor once the density profile
changes drastically.)  We see that the stable stars remain close to
their initial states for many dynamical times.  (For example, we have
evolved the hypermassive star \stard\ for twenty central rotation periods.) 
Of the stable stars, star \stara\ has density oscillations with the highest
amplitude, probably because it is very close to the TOV turning point. 
The oscillations have a period of 16.7, which is close to the expected
period for linear radial oscillations of $\sim 7\rho_c^{-1/2} = 15.7$.

\section{Black Hole Neutron Star Binary Inspiral}
\label{bhnsinspiral}

We now apply our code to model a black hole-neutron star (BHNS) binary, starting
from the late stage
of its inspiral.  To date, there have been only a few successful simulations of
these important systems in full general relativity~\cite{2007CQGra..24..125S,
Shibata:2007zm,2008PhRvD..77h4002E,Yamamoto:2008js}.  For this paper,
we have chosen to evolve an {\it equal mass} system.  Of course, we do not expect
such a system to occur frequently in nature, but we found this
choice useful for two reasons.  First, it allows us to make maximum use of our experience
with binary black holes, which has largely focused on the equal mass case.  Second, since
some of the most important work on BHNS binaries with relativistic gravity
(e.g.~\cite{2006PhRvD..73b4012F})
has been done in the limit of extreme mass ratios, an equal mass system allows us to emphasize
that we do not have this restriction.

\begin{figure}
\includegraphics[width=7.3cm]{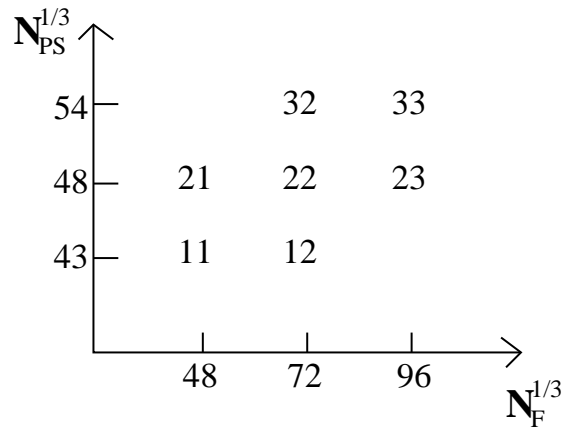}
\caption{ A representation of the choices of grid resolutions used in
  our reported runs. $N_F$ is the effective total number of grid points on the fluid
  finite difference grid, which includes ``mirror'' points given by the reflection
  symmetry about the equatorial plane; $N_{PS}$ is the total number of collocation points
  on the spectral grid.}
\label{fig:Npoints}
\end{figure}

The code used to construct our initial data is described in~\cite{Foucart:2008qt}. 
For simplicity, and to facilitate comparisons with the results of other codes, 
we have modeled the neutron star as a $\Gamma=2$ polytrope.  At the start of
the evolution, the neutron star is chosen to be irrotational, and the black hole is chosen
to have a zero quasi-local spin.  (See~\cite{Foucart:2008qt,Caudill:2006hw} for details.)  We use units
such that the mass of the black hole and the neutron star are both one.  (We define the mass
of the star to be the ADM mass of an isolated star with the same baryonic mass as the one in the
binary.  Because the black hole
has vanishing spin, we simply use the irreducible mass of the apparent
horizon as the black hole mass.)  At the
initial time, we place the black hole and neutron star in nearly circular orbit with a binary separation
of 24 in our coordinates and units.  In particular, we place the center of rotation at the
point $(0,0,0)$, the maximum density inside the star at $(-12,0,0)$, and the center of the
horizon at $(12.011,0.097,0)$.  The rotation axis is the $z$-axis, and the initial period
is $P=580$.  At this separation, the binary is expected to evolve a few orbits before the
neutron star is disrupted.  As we evolve, we use our dual frame system to hold the
center of mass of the star (which nearly coincides with its maximum density point) at
$(-12,0,0)$ in moving coordinates, and we drive the horizon center to $(12.011,0,0)$. 
After a quick adjustment, in which the black hole center moves onto $y=0$, we find that
the centers remain locked throughout the evolution.  The coordinate mapping used to track
the binary objects is a combination of two maps of type $\mathcal{RS}$ described in
Section~\ref{dual_frames}.  One $\mathcal{RS}$, labeled $\mathcal{RS}^1$, is chosen to have a
rotation axis that runs through the
origin, and its $a$ and $\Phi$ parameters are used to fix the $x$ and $y$ moving coordinates
of the center of the apparent horizon.  If the two binary objects were identical (as in the
equal mass, nonspinning binary black hole problem), this one mapping would fix both objects
by symmetry.  In our case, however, the black hole and neutron star are not identical, and
asymmetries may develop.  So we introduce a second mapping.  The second $\mathcal{RS}$, labeled
$\mathcal{RS}^2$ is chosen
to have a rotation axis at $x=12.011$, $y=0$, the location of the black hole.  This map does
not affect the location of the black hole, but its $a$ and $\Phi$ are adjusted to fix the
neutron star center of mass, defined as $\int x^i D d^3x / \int D d^3x$.  In the case reported
here, the second map is nearly an identity and so is not really needed.

\begin{figure}
\includegraphics[width=7.3cm]{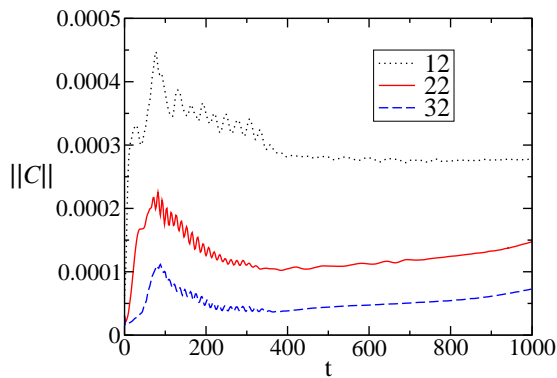}
\caption{ The normalized constraint energy $\|\mathcal{C}\|$
  defined by Eq.~(71) in~\cite{Lindblom:2005qh}. In this plot and all plots below,
  we use units in which the initial irreducible mass of the black hole is unity.}
\label{fig:GhCe}
\end{figure}

\begin{figure}
\includegraphics[width=7.3cm]{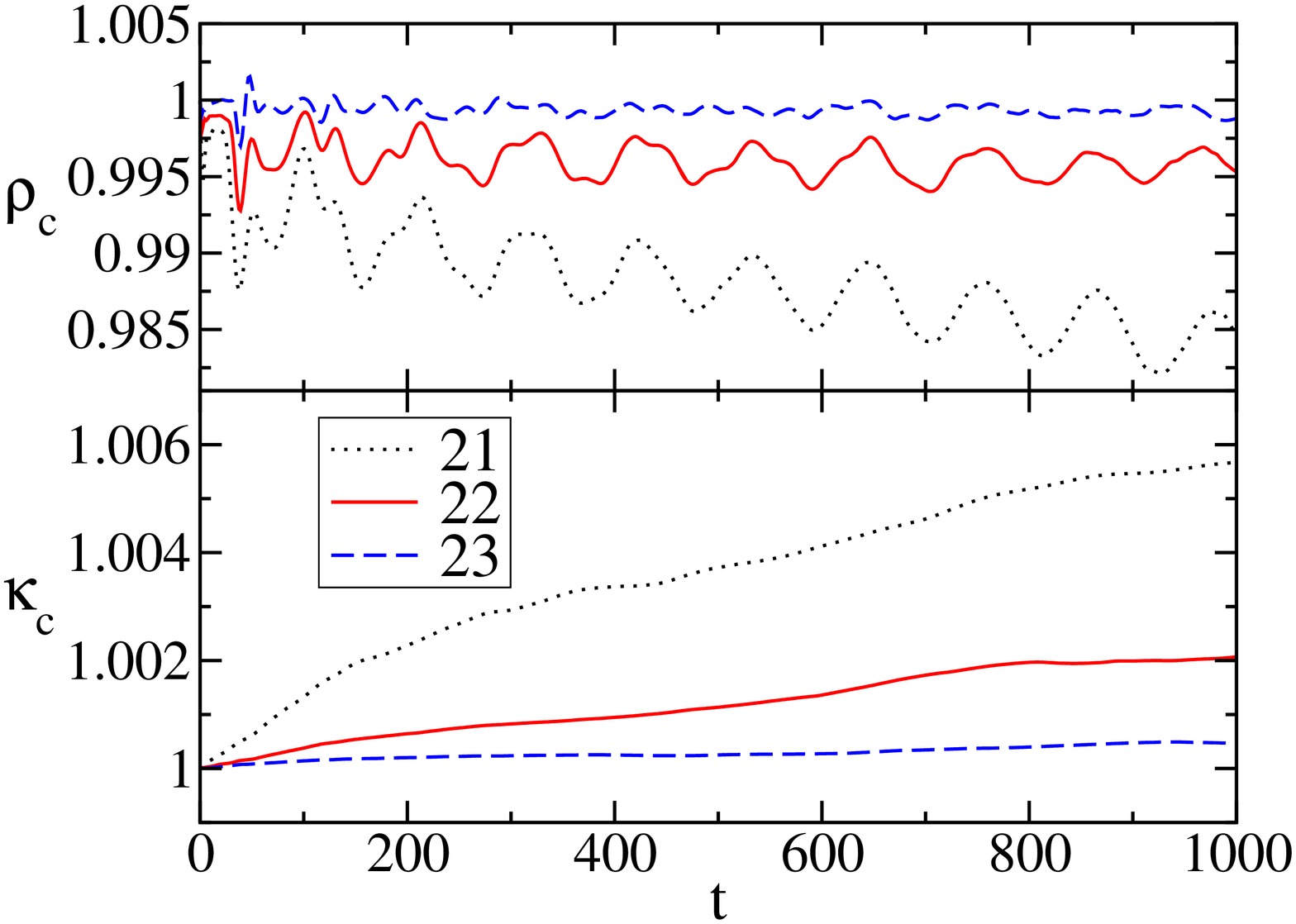}
\caption{ The baryonic density and polytropic constant
  of the neutron star, normalized
  to their initial values. }
\label{fig:density}
\end{figure}

The fluid grid is chosen to be a cube of side length 16.  (The diameter of the star is 11.5.) 
The pseudospectral grid consists of 7 cubes, 22 shells, 28 cylinders, and one cubed sphere. 
It extends to an outer radius of 326.  
In order to test the accuracy and convergence of the inspiral simulation, we vary the
resolutions of both grids as shown in Fig.~\ref{fig:Npoints}.  We evolve with three fluid grids,
having
$48^3\times {1\over 2}$, $72^3\times {1\over 2}$, $96^3\times {1\over 2}$ total gridpoints. 
(The ${1\over 2}$ factors come from our use of equatorial symmetry to evolve only above the
equator.)  We also use three PS grids, with $43^3$, $48^3$, and $54^3$ collocation
points.  The $43^3$ and the $48^3$ grids (and also the $48^3$ and $54^3$ grids) differ by
the addition of one collocation point (i.e. one new basis function) in each direction on each
domain, with the exception that the azimuthal direction on cylinders is incremented by two
points (which corresponds to adding a new sine and cosine basis function), as is needed to see convergence. 
To test boundary effects, the middle resolution (22 in Fig.~\ref{fig:Npoints}) is also run on a
pseudospectral grid with outer radius of 400. 

\begin{figure}
\includegraphics[width=7.3cm]{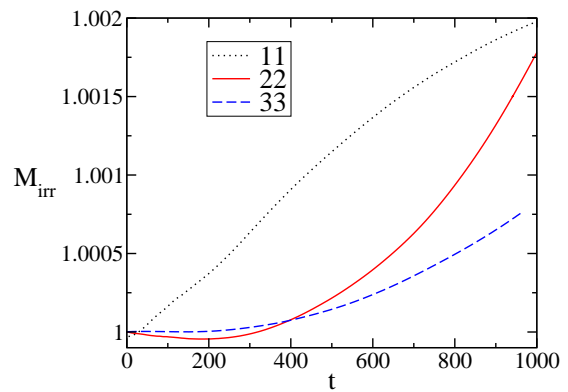}
\caption{ The irreducible mass of the black hole. }
\label{fig:BH}
\end{figure}

Each resolution is run on 32 processors on the Caltech SHC cluster.  For resolution 22,
it takes 37 hours (1184 CPU hours) to evolve one orbit at the initial separation.  At
resolution 33, it takes 100 hours (3200 CPU hours) to go one orbit.

\begin{figure}
\includegraphics[width=7.3cm]{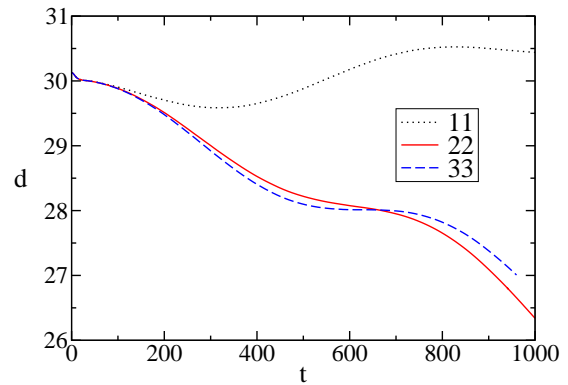}
\caption{ The proper separation between the black hole horizon
  and the neutron star center.}
\label{fig:Sep}
\end{figure}

We evolve to $t=1000$ (a little under two periods) at each resolution.  The results are
shown in figures \ref{fig:GhCe}, \ref{fig:density}, \ref{fig:BH}, and \ref{fig:Sep}.  In
Fig.~\ref{fig:GhCe}, we plot the normalized constraint energy $\|\mathcal{C}\|$
as a function of time.  We see that the
constraints converge quickly with PS resolution.  Changing the fluid grid within the
range studied has a very small effect on this diagnostic, i.e. the results for Runs 21
and 23 (not plotted) would nearly coincide with that of 22.  In the top panel of
Fig.~\ref{fig:density}, we plot the
central baryonic density normalized to its initial value.  We find that the density
converges with fluid grid resolution, but is insensitive to PS resolution over the range
studied, i.e. Runs 12 and 32 look like 22.  At the highest fluid resolution, the central
density is nearly constant, with
only a small downward drift, over the roughly two orbits of evolution shown.  Most of the
unphysical density change is caused by spurious heating.  This heating can be measured
by the effective polytropic constant $\kappa = P/\rho^{\Gamma}$.  Since there are no
physical shocks during this part of the evolution, $\kappa$ should be exactly constant. 
We plot $\kappa$ at the center of the star, normalized to its initial value, in the bottom panel of
Fig~\ref{fig:density}.  We find that the
amount of spurious heating is very small, giving us confidence in our fluid evolutions.

In Fig~\ref{fig:BH}, we plot $M_{\rm irr}$, the irreducible mass of the black hole.  Perhaps
surprisingly, we find that the error in $M_{\rm irr}$ is sensitive to both grids, so
to see convergence it is necessary to increase the resolutions of both grids simultaneously. 
Therefore, we plot $M_{\rm irr}$ for Runs 11, 22, and 33.  We see that most of the increase
in black hole mass converges away.  In Fig~\ref{fig:Sep}, we plot the proper separation $d$
between
the neutron star's center of mass and the point on the horizon on the $x$-axis facing
the star.  Once again, errors from both grids contribute, so one must use Runs 11, 22, and 33
to see convergence.  We see that even our runs at higher resolution contain some eccentricity,
presumably an artifact of the approximations, particularly the choice of zero radial velocity,
used to construct the initial data.  (See~\cite{Foucart:2008qt}, in which we remove most of this
eccentricity by adding an initial radial infall.)

\begin{figure}
\includegraphics[width=7.3cm]{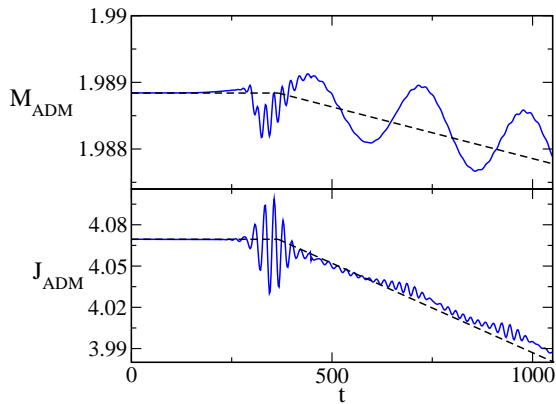}
\caption{ The mass and angular momentum of the binary.  Solid
  lines are ADM surface integrals.  Dashed lines show the expected
  changes due to losses by the observed gravitational waves.}
\label{fig:MandJ}
\end{figure}

In Fig.~\ref{fig:MandJ}, we plot the total ADM mass $M_{\rm ADM}$ and
angular momentum $J_{\rm ADM}$ of
the system as a function of time, as measured by surface integrals carried
out on the surface $r=300$.  We also monitor the gravitational wave emission
via the Newman-Penrose scalar $\psi_4$.  (See~\cite{Pfeiffer:2007yz}, for details on how
$\psi_4$ is extracted from the evolution data.)  After an initial burst of ``junk''
radiation, the true gravitational wave signal can be observed.   Nearly all
of the flux is carried in the $l=2$, $m=\pm 2$ modes.  The wave signal
is a sinusoid whose amplitude and frequency are nearly constant over
the course of our evolution.  Reading off the amplitude, frequency,
and time of arrival from the $\psi_4$ measurements, we compute the
predicted rate of mass and angular momentum loss.  These are compared with
the actual rates in Fig.~\ref{fig:MandJ}, and we see that they agree well. 
The energy flux from a point mass binary with $M=2$ and $d=24$ is
$\dot{M} = 1.61\times 10^{-6}$.  The numerical flux is
$\dot{M} = 1.55\times 10^{-6}$, so our numerical gravitational
waves are reasonable.  To test the effects of the outer boundary,
we reran the 22 case with outer radius at 420 and found no significant
change in $M_{\rm ADM}$, $J_{\rm ADM}$, or $r\psi_4$.  Also, we have varied
the extraction radii and found very small changes (apart from retardation effects)
in $M_{\rm ADM}$, $J_{\rm ADM}$, and $r\psi_4$ for $r$ between 150 and 400.

\section{Black Hole Neutron Star Binary Merger}
\label{bhnsmerger}

By continuing the evolution of the BHNS system, our code can simulate
the merger of the two objects and determine the final stationary state. 
However, we find that we must introduce several modifications to our
evolution techniques during the merger phase, which we find begins
after about four orbits of inspiral, at around $t = t_{\rm disr} \equiv 1700$.

The most obvious adjustment to be made is in the extents of the fluid grid.  When the neutron star
starts to disrupt, it is no longer possible to confine this grid to a small
box around the star.  So we must regrid at certain times to keep the matter on the
grid.  Actually, regridding may also be necessary from time to time during long inspirals,
because the neutron star grows in the moving coordinates as the scaling parameter $a$ in
the $\mathcal{RS}^1$ mapping decreases.  In our simulation, we need to expand the fluid grid
three times, at $t=1080$, $t=1540$, and $t=t_{\rm disr}$.  The final fluid grid must cover a region
around the neutron star and the black hole.  Thus, some of the savings of
our two-grid approach is lost.  Even so, the fluid grid does not need to
extend all the way into the wave zone, as it does in pure finite difference
codes.

Crucial adjustments must also be made to the dual frame control system. 
During the disruption, it is no longer appropriate to fix the center of the
neutron star.  In a sense, the BHNS problem is easier for our code than the
binary black hole problem.  In the binary black hole case, the presence of two
excision zones forces us to fix the locations of both holes in moving coordinates,
leading to an very distorted moving coordinate system as the two holes approach
each other in inertial coordinates.  In the BHNS case, we can ``release'' the
neutron star as it is disrupted, and allow it to fall into the black hole.  This
is done by setting $\Phi(t)=\Phi(t_{\rm disr})$ and $a(t)=a(t_{\rm disr})$ for
$t>t_{\rm disr}$ in the $\mathcal{RS}^2$ mapping used to fix the star.

\begin{figure}
\includegraphics[width=7.3cm]{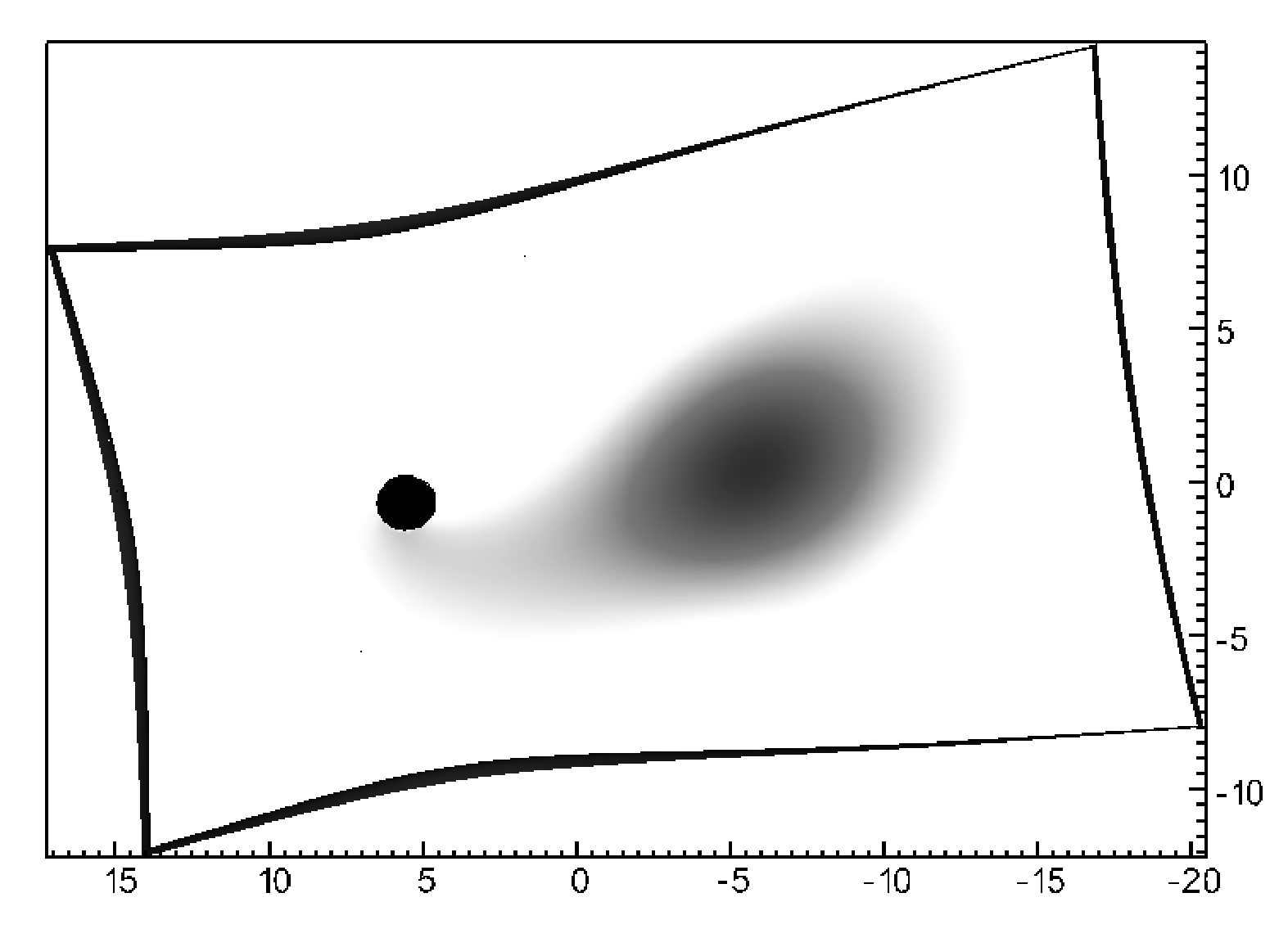} \\
\includegraphics[width=7.3cm]{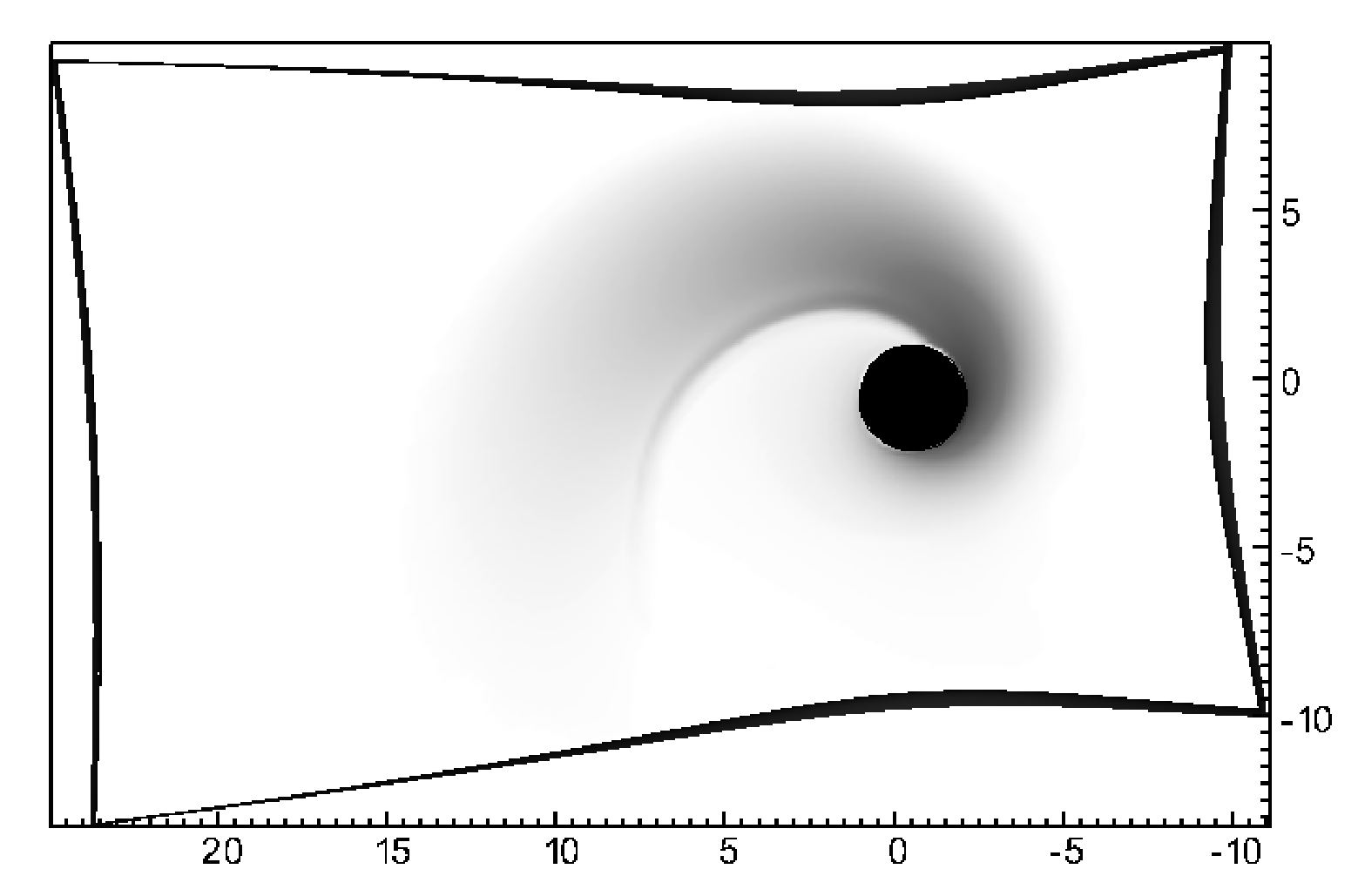} \\
\includegraphics[width=7.3cm]{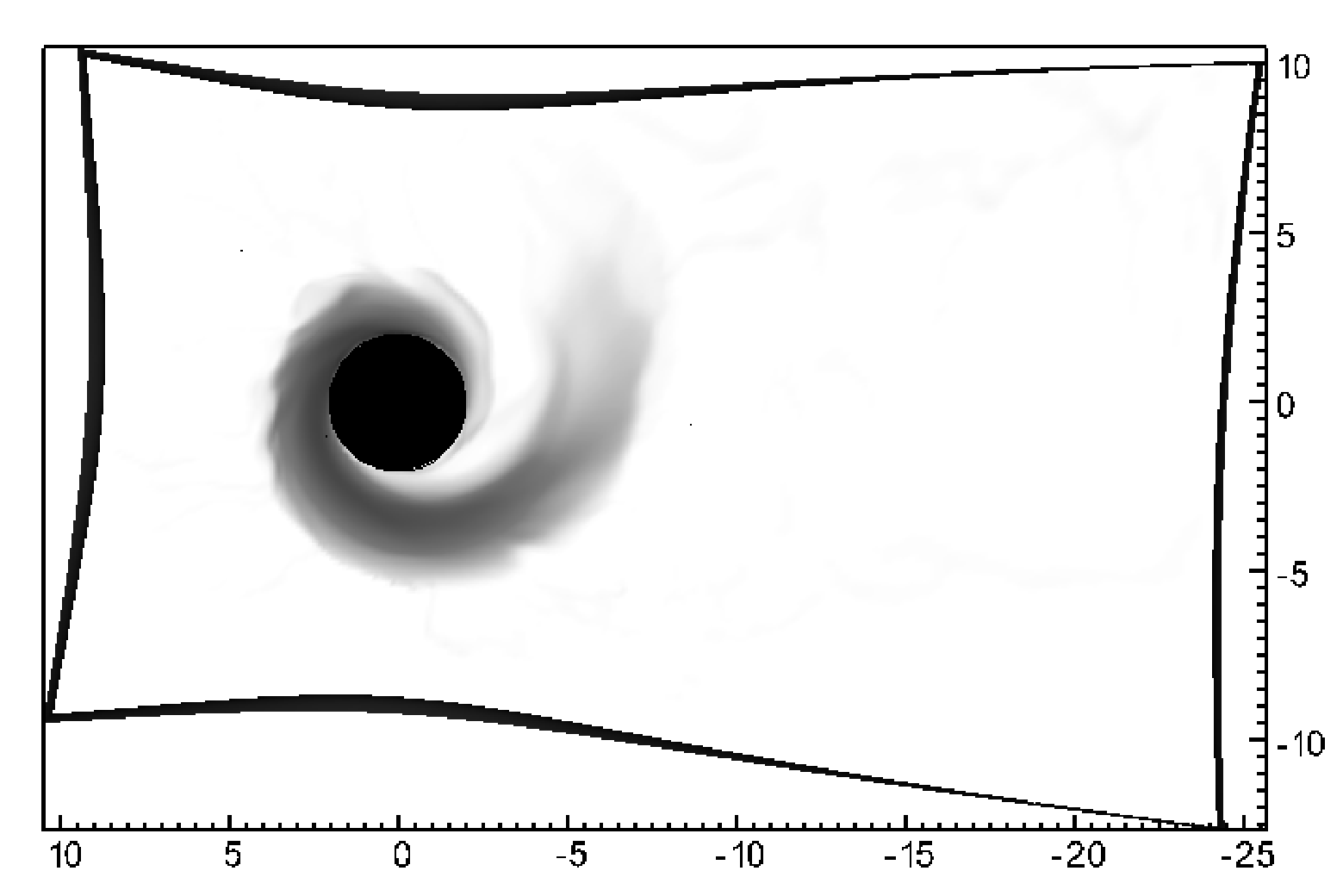}
\caption{ Snapshots of the density and horizon shape on the
  equatorial plane at times
  $t=1730$, $t=1850$, and $t=1970$.  Shades of grey represent
  $\rho/\rho_{\rm max}$, the density relative to its current
  maximum.  The black object is the apparent horizon.  The
  density and horizon are shown in inertial coordinates.  The
  distorted rectangle is the outer edge of the fluid grid, which
  is a fixed rectangle in moving coordinates but moves in the
  inertial frame.
}
\label{fig:mergerpics}
\end{figure}

Because we use excision, we must continue to fix the black hole's location on our grid. 
However, if we were to try do do this with the $\mathcal{RS}^1$ mapping, the parameter $a$
would approach zero as the hole moved towards the origin.  Instead, we switch our
coordinate control system at $t=t_{\rm disr}$.  We cease evolving the $\mathcal{RS}^1$
parameters, fixing $a(t)=a(t_{\rm disr})$ and
$\Phi(t) = \Phi(t_{\rm disr}) + \dot{\Phi}(t_{\rm disr})(t-t_{\rm disr})$
for $t>t_{\rm disr}$.  (Letting the time derivative of $\Phi$ be
continuous makes the coordinate evolution smoother.)  We then compose this mapping with another,
a simple translation:  $x^i = x^{\bar{\imath}} + C^i$.  $C^x$ and $C^y$ are used to
fix the center of the horizon.

In addition to moving, the horizon changes in size and shape, becoming much larger as it accretes
matter and becoming very distorted during the merger.  We find that our code becomes less accurate
and less stable the deeper the PS grid extends inside the hole.  Therefore, it is important to
place the excision boundary as close to the apparent horizon as possible.  We do this by introducing
another coordinate mapping that controls the size and shape of the horizon.  This mapping has
the form
\begin{equation}
\label{controlsphere}
  {\overline r} = r - f(r)\sum_{lm}\lambda_{lm}Y_{lm}(\theta,\phi)\ ,
\end{equation}
where $r$ is the coordinate distance from the horizon center $(12.011,0,0)$, $\lambda_{lm}$
are functions of time, and $f(r)$ is a smooth function of $r$.  (We use a Gaussian.) 
The $\lambda_{lm}$ parameters are used to drive the corresponding moment of the
horizon's shape in moving coordinates.  For the simulation described below, the
sum in Eq.~(\ref{controlsphere}) extends from $l=0$ to $l=6$ modes, omitting $l=1$
modes, which are controlled by the translation.  At $t=t_{\rm disr}$, we set $\lambda_{lm}=0$
for all $l>0$.  $\lambda_{00}$ is set to a value that doubles the diameter of the horizon in moving
coordinates, effectively increasing the grid coverage near the black hole.  Because
this is a discontinuous change of coordinates, we must interpolate the evolution variables
onto the new grid at $t=t_{\rm disr}$.

\begin{figure}
  \includegraphics[width=7.3cm]{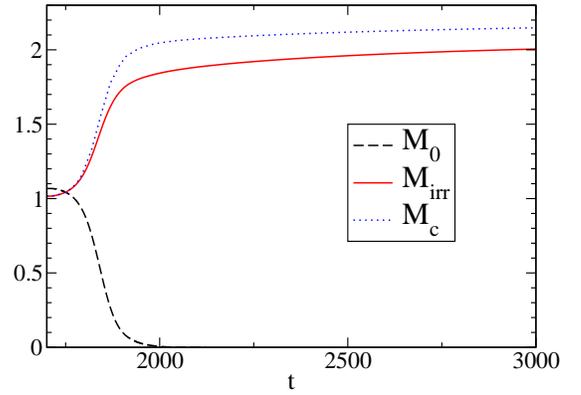}
\caption{ The evolution of the neutron star and black hole
  masses from the beginning of the star's disruption to the
  final stationary state.  Here, $M_0$ is the rest mass of
  the neutron star matter outside the hole, $M_{\rm irr}$
  is the black hole's irreducible mass, and $M_{\rm c}$ is
  the Christodoulou mass of the black hole, defined as
  $\sqrt{M_{\rm irr}^2 + J^2/(4M_{\rm irr}^2)}$, where
  $J$ is the black hole's spin. }
\label{fig:mergermasses}
\end{figure}

Finally, we find minor improvements by adjusting the generalized harmonic gauge functions $H_{a}$
during the merger.  In the future, we hope to do this using dynamical gauge conditions.  For now,
we simply damp them to zero by setting
$H_{a}(t) = H_{a}(0)e^{-((t-t_{\rm disr})/\tau)^2}$, where
$\tau=100$.  Our evolutions seem to be fairly insensitive to $\tau$; it can be varied by an order
of magnitude with no significant effects.

\begin{figure}
\includegraphics[width=7.3cm]{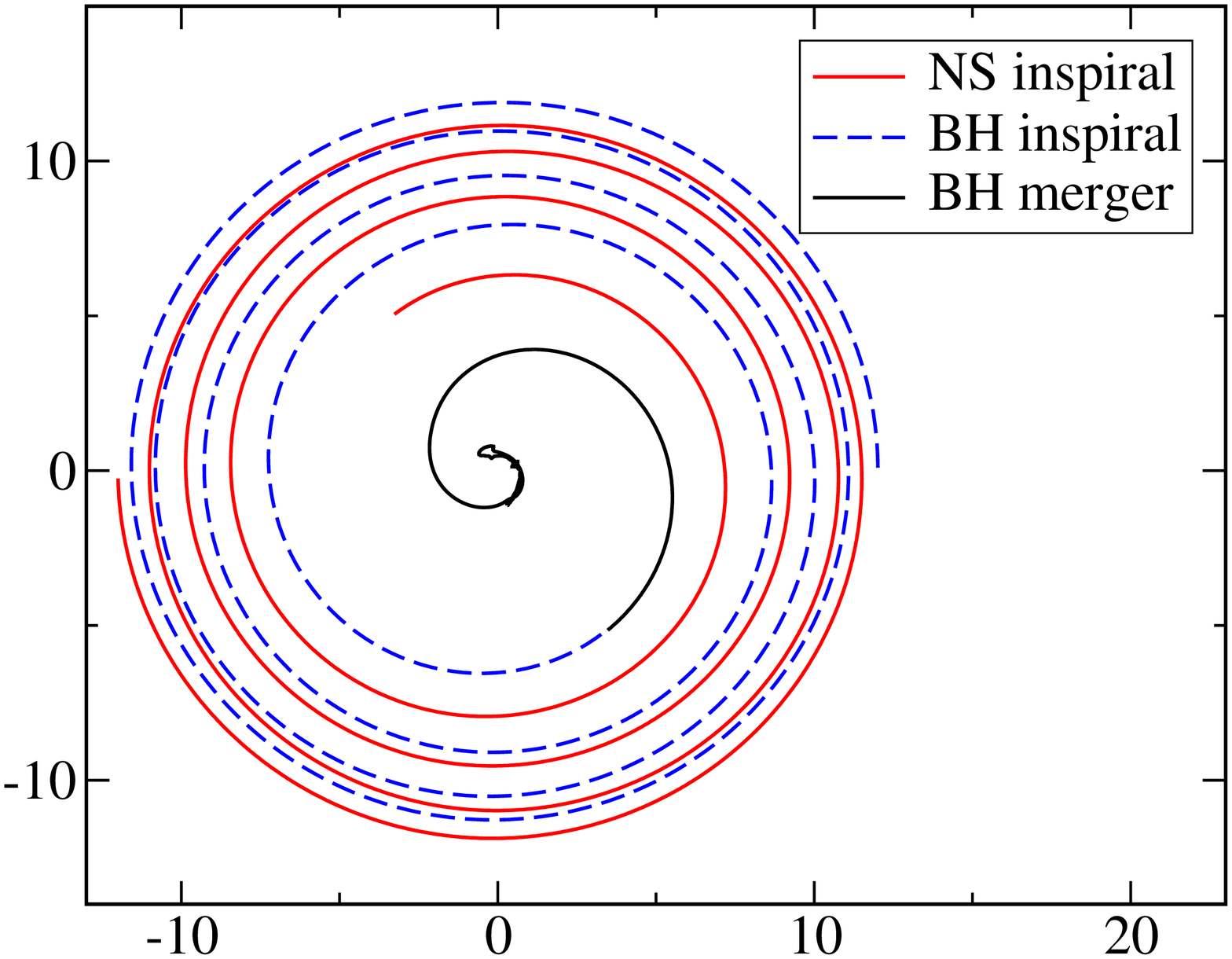}
\caption{ The trajectory in inertial coordinates of the neutron star center from $t=0$ to $t=t_{\rm disr}$,
  and that of the black hole horizon center from $t=0$ to $t=3000$.  The black hole line switches from
  dashed to solid at $t=t_{\rm disr}$.  Both the black hole and the neutron star remain on the equatorial
  plane throughout the evolution.
}
\label{fig:trajectory}
\end{figure}

For our merger simulation, we begin at $t=1000$ with the inspiral data computed on the grids
corresponding to resolution 22 and PS outer radius of 400.  We then continue the evolution
of this system to later times.  The evolution
from $t=1000$ to $t=1700$ is very similar to the evolution from $t=0$ to $t=1000$ described
in Section~\ref{bhnsinspiral}.  At $t=1700$, matter is starting to flow off the neutron star,
so we introduce the changes described above, continuing
the evolution on 80 processors in order to accommodate the larger fluid grid needed.  The
resolution on the final fluid grid is $\Delta x = 0.4$, so that there are 20 fluid grid points
across the diameter of the excision zone.  On the LONI Queen Bee cluster, the code must run
for 43 hours (3440 CPU hours) to evolve from $t=1700$ to $t=2000$, by which time most of the matter
has fallen into the hole.

Three snapshots of the neutron star density are shown in Fig.~\ref{fig:mergerpics}. 
A dense stream of matter flows from the star to the black hole, so that there quickly
come to be two peaks in the matter density:  one at the center of the neutron star and
the other at the point where the stream reaches the hole. 
  The neutron star continues to fall closer to the hole, even as it rapidly loses mass
through the matter stream.  The density peak corresponding to the neutron star core
disappears at around $t=1800$.  At around $t=1850$, the matter stream starts to close
in a ring around the black hole.  A shock forms at this time where matter
flowing towards the hole intersects matter flowing around the hole.  This can be seen
in the sharp inner edge of the grey swath in the middle panel of Fig.~\ref{fig:mergerpics}. 
Matter falls rapidly into the hole from $t=1750$ until $t=2000$.

\begin{figure}
\includegraphics[width=7.3cm]{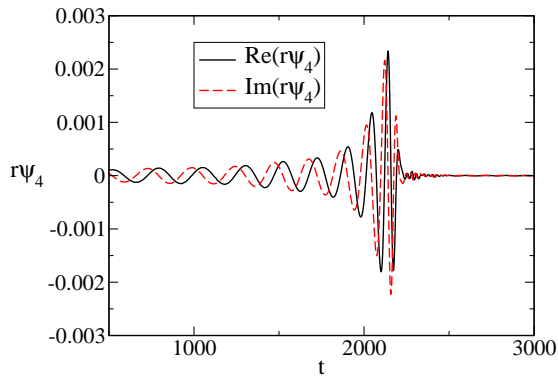}
\caption{ The gravitational wave signal for the entire inspiral
  and merger (excluding the initial burst of ``junk'' radiation). The
  quadrupole contribution dominates, so we plot only the $l=m=2$ amplitudes.
}
\label{fig:mergerpsi4}
\end{figure}

In Fig.~\ref{fig:mergermasses}, we plot the baryon rest mass $M_0$ on the grid together with the
apparent horizon irreducible mass $M_{\rm irr}$ as functions of time starting from $t=t_{\rm disr}$. 
We find that virtually all of the matter falls promptly into the black hole; $M_0$ drops
from its initial value near unity until it stabilizes about $2\times 10^{-4}$, indicating a very
small post-merger accretion disk. 
This result is consistent with the small disks found by other groups (particularly
Etienne~{\it et al}~\cite{2008PhRvD..77h4002E}, who have
also simulated the equal mass case, but also the most recent results
of Yamamoto, Shibata, and Taniguchi~\cite{Yamamoto:2008js}).  At $t=2200$, the matter is so
sparse that we stop evolving the fluid and continue the metric evolution using our vacuum Einstein code. 

The final $M_{\rm irr}$ of the black hole is very close to 2, and its final spin is $J/M^2\approx 0.7$. 
(The computations used to obtain the black hole spin are described in the appendicies
of~\cite{Lovelace:2008tw}.) 
From $M_{\rm irr}$ and $J$, we can compute the Christodoulou mass, and it is found to be
$M_{\rm c}\approx 2.15$.  That $M_{\rm c}$ is greater than 2 is presumably a result of the
error in our calculation and is consistent with the observed constraint violation (see below). 
The complete trajectories in inertial coordinates of the black hole horizon center and the neutron
star center of mass are plotted in~Fig.~\ref{fig:trajectory}. 
In Fig.~\ref{fig:mergerpsi4}, we plot the $l=m=2$ component of $r\psi_4$ for the complete evolution,
extracting the wave at $r=350$.

\begin{figure}
\includegraphics[width=7.3cm]{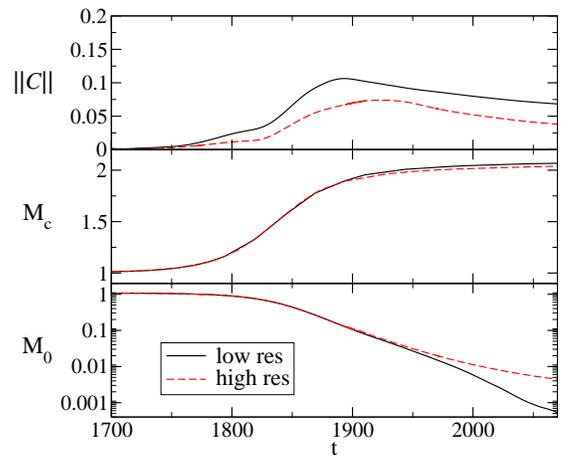}
\caption{ The constraint energy $\|\mathcal{C}\|$, black hole mass $M_{\rm c}$, and
  baryonic rest mass $M_0$ for the merger phase in evolutions carried out at two
  different resolutions.
}
\label{fig:mergerconvg}
\end{figure}

To test the stability and convergence rate of our merger algorithm, we have run the merger phase
$1700 < t < 2050$ at a higher resolution.  For this run, we increase the number of points on the
fluid grid by 40\% and the number of PS collocation points by one per axis per domain, keeping
the extents of both grids fixed.  The code runs about four times slower on the larger grid. The
effects of this grid change are shown in
Fig.~\ref{fig:mergerconvg}.  The constraint violation peaks at $t\approx 1900$ at about
$\|\mathcal{C}\|=0.1$ for the low resolution run and at around $\|\mathcal{C}\|=0.07$ for the
high resolution run.  The constraint violations decline thereafter.  We note that, since the
constraint energy is strongly peaked near the black hole, these numbers depend sensitively on
how the norm is taken.  The quantity $\|\mathcal{C}\|$ is an L2-type norm of the form
$\left(\int f^2 dV\right)^{1/2}$, where $f$ represents the generalized harmonic constraints. 
(See Eq.~53 and Eq.~71 of~\cite{Lindblom:2005qh}.)  If we instead use an L1-type norm of the
form $\int \left|f\right| dV$, we find that this norm, when appropriately normalized, peaks at
0.02 for the low resolution run and 0.01 for the high resolution run.  The Hamiltonian and momentum
constraints (integrated with L1-type norms) show similar levels of violation.  The relative momentum
constraint violation peaks at about 0.02 for the low resolution run and 0.01 for the high resolution
run.  The relative Hamiltonian constraint violation is about 0.02 for both resolutions---convervence
is not seen because this error is present in the data at $t=t_{\rm disr}$ from which both runs start. 
Of course, these numbers are also very sensitive to the way the constraints are summed and normalized. 
The post-merger black hole mass is slightly lower at higher
resolution.  The accretion rate is also lower, but only after about 98\% of the rest mass has already
fallen into the hole.  In neither run do we find a long-lived massive disk.

%Finite difference codes have achieved Hamiltonian constraint violations of around 5\% (see, e.g. Fig.~5
%in~\cite{2007CQGra..24..125S} and Fig.~2 in~\cite{2008PhRvD..77h4002E}).  However, it may be deceptive
%to directly compare these numbers, because the constraint violation integrals used in these other papers
%are of the form $\int \left|f\right| dV$ (see Eq.~43 of~\cite{2008PhRvD..77h4002E} and Eq.~42
%of~\cite{PhysRevD.67.024033}), where $f$ is some constraint, while $\|\mathcal{C}\|$ is a norm of
%the form $\left(\int f^2 dV\right)^{1/2}$ (see Eq.~53 and Eq.~71 of~\cite{Lindblom:2005qh}).  If we
%integrate the absolute values of the generalized harmonic constraint violations, we find a relative
%error of 0.015.  Integrating the Hamiltonian constraint violation in the same way, we find a relative
%error of 0.02.  We have evolved this system from $t=1700$ to $t=1960$ using a denser PS grid and a denser
%fluid grid, and we confirmed that the constraint violation does decrease with increasing resolution.

\section{Conclusions}
\label{conclusions}

We have shown that our code can evolve inspiraling BHNS binaries with high
accuracy at fairly low computational cost.  This enables us to begin
simulations at relatively large binary separations.  Long inspirals may
turn out to be very important for accurately modeling mergers; merger
simulations by the Illinois group show some sensitivity to the initial
separation~\cite{2008PhRvD..77h4002E}, with the implication that starting too close to the
merger can lead to underestimating the mass of the post-merger
accretion disk.  We have also shown that we can simulate the merger of
these binaries, although our accuracy of these simulations is not as good
so far than that of our inspirals.  Also, there is every reason to believe that the
techniques described here would work just as well for binary neutron stars.

An important next step is to demonstrate that our code can evolve more
general BHNS binaries.  We have recently begun evolving such systems with different
mass ratios and black hole spins, and so far we have had little difficulty in
evolving the inspirals.  We hope to report on these simulations in the near future. 
Also, we plan to study ways to improve the accuracy of our merger simulations.

It has only been two years since the first fully relativistic BHNS merger
simulations were reported, and as yet very little of the parameter
space has been studied.  In particular, there is a pressing need to study
the effects of the black hole spin and the neutron star equation of state. 
There are preliminary indications (e.g.~\cite{Rantsiou:2007ct,Shibata:2007zm})
that both of these have important
effects on the post-merger accretion disk mass and the gravitational wave signal. 
Also, both of the other
groups currently performing BHNS merger simulations use very
similar techniques (BSSN, moving punctures).  It will be useful to compare
their results with those obtained using very different techniques, like
the ones reported here.

\acknowledgments

We thank Francois Limousin and Manuel Tiglio for useful discussions.  This
work was supported in part by a grant from the Sherman Fairchild Foundation,
by NSF grants PHY-0652952, DMS-0553677, PHY-0652929, and
NASA grant NNG05GG51G.  This research was supported in part by the NSF
through TeraGrid~\cite{teragrid} resources provided by LONI's Queen Bee cluster. 
Computations were also performed on Caltech's Shared Heterogeneous Cluster (SHC).

%\appendix

%\section{}
%\label{}

\bibliography{References/References}

\begin{thebibliography}{75}
\expandafter\ifx\csname natexlab\endcsname\relax\def\natexlab#1{#1}\fi
\expandafter\ifx\csname bibnamefont\endcsname\relax
  \def\bibnamefont#1{#1}\fi
\expandafter\ifx\csname bibfnamefont\endcsname\relax
  \def\bibfnamefont#1{#1}\fi
\expandafter\ifx\csname citenamefont\endcsname\relax
  \def\citenamefont#1{#1}\fi
\expandafter\ifx\csname url\endcsname\relax
  \def\url#1{\texttt{#1}}\fi
\expandafter\ifx\csname urlprefix\endcsname\relax\def\urlprefix{URL }\fi
\providecommand{\bibinfo}[2]{#2}
\providecommand{\eprint}[2][]{\url{#2}}

\bibitem[{\citenamefont{{Sadowski} et~al.}(2008)\citenamefont{{Sadowski},
  {Belczynski}, {Bulik}, {Ivanova}, {Rasio}, and
  {O'Shaughnessy}}}]{2008ApJ...676.1162S}
\bibinfo{author}{\bibfnamefont{A.}~\bibnamefont{{Sadowski}}},
  \bibinfo{author}{\bibfnamefont{K.}~\bibnamefont{{Belczynski}}},
  \bibinfo{author}{\bibfnamefont{T.}~\bibnamefont{{Bulik}}},
  \bibinfo{author}{\bibfnamefont{N.}~\bibnamefont{{Ivanova}}},
  \bibinfo{author}{\bibfnamefont{F.~A.} \bibnamefont{{Rasio}}},
  \bibnamefont{and}
  \bibinfo{author}{\bibfnamefont{R.}~\bibnamefont{{O'Shaughnessy}}},
  \bibinfo{journal}{\apj} \textbf{\bibinfo{volume}{676}}, \bibinfo{pages}{1162}
  (\bibinfo{year}{2008}), \eprint{astro-ph/0710.0878}.

\bibitem[{\citenamefont{{Narayan} et~al.}(1992)\citenamefont{{Narayan},
  {Paczynski}, and {Piran}}}]{1992ApJ...395L..83N}
\bibinfo{author}{\bibfnamefont{R.}~\bibnamefont{{Narayan}}},
  \bibinfo{author}{\bibfnamefont{B.}~\bibnamefont{{Paczynski}}},
  \bibnamefont{and} \bibinfo{author}{\bibfnamefont{T.}~\bibnamefont{{Piran}}},
  \bibinfo{journal}{Astrophys.\ J.\ Letters} \textbf{\bibinfo{volume}{395}},
  \bibinfo{pages}{L83} (\bibinfo{year}{1992}), \eprint{astro-ph/9204001}.

\bibitem[{\citenamefont{{Gehrels} et~al.}(2005)\citenamefont{{Gehrels},
  {Sarazin}, {O'Brien}, {Zhang}, {Barbier}, {Barthelmy}, {Blustin}, {Burrows},
  {Cannizzo}, {Cummings} et~al.}}]{2005Natur.437..851G}
\bibinfo{author}{\bibfnamefont{N.}~\bibnamefont{{Gehrels}}},
  \bibinfo{author}{\bibfnamefont{C.~L.} \bibnamefont{{Sarazin}}},
  \bibinfo{author}{\bibfnamefont{P.~T.} \bibnamefont{{O'Brien}}},
  \bibinfo{author}{\bibfnamefont{B.}~\bibnamefont{{Zhang}}},
  \bibinfo{author}{\bibfnamefont{L.}~\bibnamefont{{Barbier}}},
  \bibinfo{author}{\bibfnamefont{S.~D.} \bibnamefont{{Barthelmy}}},
  \bibinfo{author}{\bibfnamefont{A.}~\bibnamefont{{Blustin}}},
  \bibinfo{author}{\bibfnamefont{D.~N.} \bibnamefont{{Burrows}}},
  \bibinfo{author}{\bibfnamefont{J.}~\bibnamefont{{Cannizzo}}},
  \bibinfo{author}{\bibfnamefont{J.~R.} \bibnamefont{{Cummings}}},
  \bibnamefont{et~al.}, \bibinfo{journal}{\nat} \textbf{\bibinfo{volume}{437}},
  \bibinfo{pages}{851} (\bibinfo{year}{2005}), \eprint{astro-ph/0505630}.

\bibitem[{\citenamefont{{Villasenor} et~al.}(2005)\citenamefont{{Villasenor},
  {Lamb}, {Ricker}, {Atteia}, {Kawai}, {Butler}, {Nakagawa}, {Jernigan},
  {Boer}, {Crew} et~al.}}]{2005Natur.437..855V}
\bibinfo{author}{\bibfnamefont{J.~S.} \bibnamefont{{Villasenor}}},
  \bibinfo{author}{\bibfnamefont{D.~Q.} \bibnamefont{{Lamb}}},
  \bibinfo{author}{\bibfnamefont{G.~R.} \bibnamefont{{Ricker}}},
  \bibinfo{author}{\bibfnamefont{J.-L.} \bibnamefont{{Atteia}}},
  \bibinfo{author}{\bibfnamefont{N.}~\bibnamefont{{Kawai}}},
  \bibinfo{author}{\bibfnamefont{N.}~\bibnamefont{{Butler}}},
  \bibinfo{author}{\bibfnamefont{Y.}~\bibnamefont{{Nakagawa}}},
  \bibinfo{author}{\bibfnamefont{J.~G.} \bibnamefont{{Jernigan}}},
  \bibinfo{author}{\bibfnamefont{M.}~\bibnamefont{{Boer}}},
  \bibinfo{author}{\bibfnamefont{G.~B.} \bibnamefont{{Crew}}},
  \bibnamefont{et~al.}, \bibinfo{journal}{\nat} \textbf{\bibinfo{volume}{437}},
  \bibinfo{pages}{855} (\bibinfo{year}{2005}).

\bibitem[{\citenamefont{{Fox} et~al.}(2005)\citenamefont{{Fox}, {Frail},
  {Price}, {Kulkarni}, {Berger}, {Piran}, {Soderberg}, {Cenko}, {Cameron},
  {Gal-Yam} et~al.}}]{2005Natur.437..845F}
\bibinfo{author}{\bibfnamefont{D.~B.} \bibnamefont{{Fox}}},
  \bibinfo{author}{\bibfnamefont{D.~A.} \bibnamefont{{Frail}}},
  \bibinfo{author}{\bibfnamefont{P.~A.} \bibnamefont{{Price}}},
  \bibinfo{author}{\bibfnamefont{S.~R.} \bibnamefont{{Kulkarni}}},
  \bibinfo{author}{\bibfnamefont{E.}~\bibnamefont{{Berger}}},
  \bibinfo{author}{\bibfnamefont{T.}~\bibnamefont{{Piran}}},
  \bibinfo{author}{\bibfnamefont{A.~M.} \bibnamefont{{Soderberg}}},
  \bibinfo{author}{\bibfnamefont{S.~B.} \bibnamefont{{Cenko}}},
  \bibinfo{author}{\bibfnamefont{P.~B.} \bibnamefont{{Cameron}}},
  \bibinfo{author}{\bibfnamefont{A.}~\bibnamefont{{Gal-Yam}}},
  \bibnamefont{et~al.}, \bibinfo{journal}{Nature}
  \textbf{\bibinfo{volume}{437}}, \bibinfo{pages}{845} (\bibinfo{year}{2005}),
  \eprint{astro-ph/0510110}.

\bibitem[{\citenamefont{{Lattimer} and {Schramm}}(1974)}]{1974ApJ...192L.145L}
\bibinfo{author}{\bibfnamefont{J.~M.} \bibnamefont{{Lattimer}}}
  \bibnamefont{and} \bibinfo{author}{\bibfnamefont{D.~N.}
  \bibnamefont{{Schramm}}}, \bibinfo{journal}{Astrophys.\ J.\ Letters}
  \textbf{\bibinfo{volume}{192}}, \bibinfo{pages}{L145} (\bibinfo{year}{1974}).

\bibitem[{\citenamefont{{Rosswog}}(2005)}]{2005ApJ...634.1202R}
\bibinfo{author}{\bibfnamefont{S.}~\bibnamefont{{Rosswog}}},
  \bibinfo{journal}{Astrophys.\ J.} \textbf{\bibinfo{volume}{634}},
  \bibinfo{pages}{1202} (\bibinfo{year}{2005}), \eprint{astro-ph/0508138}.

\bibitem[{\citenamefont{{Miller}}(2005)}]{2005ApJ...626L..41M}
\bibinfo{author}{\bibfnamefont{M.~C.} \bibnamefont{{Miller}}},
  \bibinfo{journal}{Astrophys.\ J.\ Letters} \textbf{\bibinfo{volume}{626}},
  \bibinfo{pages}{L41} (\bibinfo{year}{2005}), \eprint{astro-ph/0505094}.

\bibitem[{\citenamefont{Rantsiou et~al.}(2007)\citenamefont{Rantsiou,
  Kobayashi, Laguna, and Rasio}}]{Rantsiou:2007ct}
\bibinfo{author}{\bibfnamefont{E.}~\bibnamefont{Rantsiou}},
  \bibinfo{author}{\bibfnamefont{S.}~\bibnamefont{Kobayashi}},
  \bibinfo{author}{\bibfnamefont{P.}~\bibnamefont{Laguna}}, \bibnamefont{and}
  \bibinfo{author}{\bibfnamefont{F.}~\bibnamefont{Rasio}}
  (\bibinfo{year}{2007}), \eprint{astro-ph/0703599}.

\bibitem[{\citenamefont{{Ruffert} et~al.}(1996)\citenamefont{{Ruffert},
  {Janka}, and {Schaefer}}}]{1996A&A...311..532R}
\bibinfo{author}{\bibfnamefont{M.}~\bibnamefont{{Ruffert}}},
  \bibinfo{author}{\bibfnamefont{H.-T.} \bibnamefont{{Janka}}},
  \bibnamefont{and}
  \bibinfo{author}{\bibfnamefont{G.}~\bibnamefont{{Schaefer}}},
  \bibinfo{journal}{Astron.\ and Astrophys.} \textbf{\bibinfo{volume}{311}},
  \bibinfo{pages}{532} (\bibinfo{year}{1996}), \eprint{astro-ph/9509006}.

\bibitem[{\citenamefont{{Ruffert} et~al.}(1997)\citenamefont{{Ruffert},
  {Janka}, {Takahashi}, and {Schaefer}}}]{1997A&A...319..122R}
\bibinfo{author}{\bibfnamefont{M.}~\bibnamefont{{Ruffert}}},
  \bibinfo{author}{\bibfnamefont{H.-T.} \bibnamefont{{Janka}}},
  \bibinfo{author}{\bibfnamefont{K.}~\bibnamefont{{Takahashi}}},
  \bibnamefont{and}
  \bibinfo{author}{\bibfnamefont{G.}~\bibnamefont{{Schaefer}}},
  \bibinfo{journal}{Astron.\ and Astrophys.} \textbf{\bibinfo{volume}{319}},
  \bibinfo{pages}{122} (\bibinfo{year}{1997}), \eprint{astro-ph/9606181}.

\bibitem[{\citenamefont{{Rosswog} et~al.}(1999)\citenamefont{{Rosswog},
  {Liebend{\"o}rfer}, {Thielemann}, {Davies}, {Benz}, and
  {Piran}}}]{1999A&A...341..499R}
\bibinfo{author}{\bibfnamefont{S.}~\bibnamefont{{Rosswog}}},
  \bibinfo{author}{\bibfnamefont{M.}~\bibnamefont{{Liebend{\"o}rfer}}},
  \bibinfo{author}{\bibfnamefont{F.-K.} \bibnamefont{{Thielemann}}},
  \bibinfo{author}{\bibfnamefont{M.~B.} \bibnamefont{{Davies}}},
  \bibinfo{author}{\bibfnamefont{W.}~\bibnamefont{{Benz}}}, \bibnamefont{and}
  \bibinfo{author}{\bibfnamefont{T.}~\bibnamefont{{Piran}}},
  \bibinfo{journal}{Astron.\ and Astrophys.} \textbf{\bibinfo{volume}{341}},
  \bibinfo{pages}{499} (\bibinfo{year}{1999}), \eprint{astro-ph/9811367}.

\bibitem[{\citenamefont{{Rosswog} and {Davies}}(2002)}]{2002MNRAS.334..481R}
\bibinfo{author}{\bibfnamefont{S.}~\bibnamefont{{Rosswog}}} \bibnamefont{and}
  \bibinfo{author}{\bibfnamefont{M.~B.} \bibnamefont{{Davies}}},
  \bibinfo{journal}{MNRAS} \textbf{\bibinfo{volume}{334}}, \bibinfo{pages}{481}
  (\bibinfo{year}{2002}).

\bibitem[{\citenamefont{Dessart et~al.}(2008)\citenamefont{Dessart, Ott,
  Burrows, Rosswog, and Livne}}]{2008arXiv0806.4380D}
\bibinfo{author}{\bibfnamefont{L.}~\bibnamefont{Dessart}},
  \bibinfo{author}{\bibfnamefont{C.}~\bibnamefont{Ott}},
  \bibinfo{author}{\bibfnamefont{A.}~\bibnamefont{Burrows}},
  \bibinfo{author}{\bibfnamefont{S.}~\bibnamefont{Rosswog}}, \bibnamefont{and}
  \bibinfo{author}{\bibfnamefont{E.}~\bibnamefont{Livne}}
  (\bibinfo{year}{2008}), \eprint{astro-ph/0806.4380}.

\bibitem[{\citenamefont{{Price} and {Rosswog}}(2006)}]{2006Sci...312..719P}
\bibinfo{author}{\bibfnamefont{D.~J.} \bibnamefont{{Price}}} \bibnamefont{and}
  \bibinfo{author}{\bibfnamefont{S.}~\bibnamefont{{Rosswog}}},
  \bibinfo{journal}{Science} \textbf{\bibinfo{volume}{312}},
  \bibinfo{pages}{719} (\bibinfo{year}{2006}), \eprint{astro-ph/0603845}.

\bibitem[{\citenamefont{Oechslin et~al.}(2006)\citenamefont{Oechslin, Janka,
  and Marek}}]{Oechslin:2006uk}
\bibinfo{author}{\bibfnamefont{R.}~\bibnamefont{Oechslin}},
  \bibinfo{author}{\bibfnamefont{H.~T.} \bibnamefont{Janka}}, \bibnamefont{and}
  \bibinfo{author}{\bibfnamefont{A.}~\bibnamefont{Marek}}
  (\bibinfo{year}{2006}), \eprint{astro-ph/0611047}.

\bibitem[{\citenamefont{{Shibata} and {Ury{\= u}}}(2000)}]{2000PhRvD..61f4001S}
\bibinfo{author}{\bibfnamefont{M.}~\bibnamefont{{Shibata}}} \bibnamefont{and}
  \bibinfo{author}{\bibfnamefont{K.}~\bibnamefont{{Ury{\= u}}}},
  \bibinfo{journal}{Phys.\ Rev.\ D} \textbf{\bibinfo{volume}{61}},
  \bibinfo{pages}{064001} (\bibinfo{year}{2000}), \eprint{gr-qc/9911058}.

\bibitem[{\citenamefont{{Shibata} and {Ury{\= u}}}(2002)}]{2002PThPh.107..265S}
\bibinfo{author}{\bibfnamefont{M.}~\bibnamefont{{Shibata}}} \bibnamefont{and}
  \bibinfo{author}{\bibfnamefont{K.}~\bibnamefont{{Ury{\= u}}}},
  \bibinfo{journal}{Progress of Theoretical Physics}
  \textbf{\bibinfo{volume}{107}}, \bibinfo{pages}{265} (\bibinfo{year}{2002}),
  \eprint{gr-qc/0203037}.

\bibitem[{\citenamefont{{Shibata} et~al.}(2003)\citenamefont{{Shibata},
  {Taniguchi}, and {Ury{\= u}}}}]{2003PhRvD..68h4020S}
\bibinfo{author}{\bibfnamefont{M.}~\bibnamefont{{Shibata}}},
  \bibinfo{author}{\bibfnamefont{K.}~\bibnamefont{{Taniguchi}}},
  \bibnamefont{and} \bibinfo{author}{\bibfnamefont{K.}~\bibnamefont{{Ury{\=
  u}}}}, \bibinfo{journal}{Phys.\ Rev.\ D} \textbf{\bibinfo{volume}{68}},
  \bibinfo{pages}{084020} (\bibinfo{year}{2003}), \eprint{gr-qc/0310030}.

\bibitem[{\citenamefont{{Marronetti} et~al.}(2004)\citenamefont{{Marronetti},
  {Duez}, {Shapiro}, and {Baumgarte}}}]{2004PhRvL..92n1101M}
\bibinfo{author}{\bibfnamefont{P.}~\bibnamefont{{Marronetti}}},
  \bibinfo{author}{\bibfnamefont{M.~D.} \bibnamefont{{Duez}}},
  \bibinfo{author}{\bibfnamefont{S.~L.} \bibnamefont{{Shapiro}}},
  \bibnamefont{and} \bibinfo{author}{\bibfnamefont{T.~W.}
  \bibnamefont{{Baumgarte}}}, \bibinfo{journal}{Physical Review Letters}
  \textbf{\bibinfo{volume}{92}}, \bibinfo{pages}{141101}
  (\bibinfo{year}{2004}), \eprint{gr-qc/0312036}.

\bibitem[{\citenamefont{{Miller} et~al.}(2004)\citenamefont{{Miller},
  {Gressman}, and {Suen}}}]{2004PhRvD..69f4026M}
\bibinfo{author}{\bibfnamefont{M.}~\bibnamefont{{Miller}}},
  \bibinfo{author}{\bibfnamefont{P.}~\bibnamefont{{Gressman}}},
  \bibnamefont{and} \bibinfo{author}{\bibfnamefont{W.-M.}
  \bibnamefont{{Suen}}}, \bibinfo{journal}{Phys.\ Rev.\ D}
  \textbf{\bibinfo{volume}{69}}, \bibinfo{pages}{064026}
  (\bibinfo{year}{2004}), \eprint{gr-qc/0312030}.

\bibitem[{\citenamefont{Baiotti et~al.}(2008)\citenamefont{Baiotti, Giacomazzo,
  and Rezzolla}}]{Baiotti:2008ra}
\bibinfo{author}{\bibfnamefont{L.}~\bibnamefont{Baiotti}},
  \bibinfo{author}{\bibfnamefont{B.}~\bibnamefont{Giacomazzo}},
  \bibnamefont{and} \bibinfo{author}{\bibfnamefont{L.}~\bibnamefont{Rezzolla}}
  (\bibinfo{year}{2008}), \eprint{gr-qc/0804.0594}.

\bibitem[{\citenamefont{{Shibata} et~al.}(2005)\citenamefont{{Shibata},
  {Taniguchi}, and {Ury{\= u}}}}]{2005PhRvD..71h4021S}
\bibinfo{author}{\bibfnamefont{M.}~\bibnamefont{{Shibata}}},
  \bibinfo{author}{\bibfnamefont{K.}~\bibnamefont{{Taniguchi}}},
  \bibnamefont{and} \bibinfo{author}{\bibfnamefont{K.}~\bibnamefont{{Ury{\=
  u}}}}, \bibinfo{journal}{Phys.\ Rev.\ D} \textbf{\bibinfo{volume}{71}},
  \bibinfo{pages}{084021} (\bibinfo{year}{2005}), \eprint{gr-qc/0503119}.

\bibitem[{\citenamefont{Shibata and Taniguchi}(2006)}]{Shibata:2006nm}
\bibinfo{author}{\bibfnamefont{M.}~\bibnamefont{Shibata}} \bibnamefont{and}
  \bibinfo{author}{\bibfnamefont{K.}~\bibnamefont{Taniguchi}},
  \bibinfo{journal}{Phys. Rev.} \textbf{\bibinfo{volume}{D73}},
  \bibinfo{pages}{064027} (\bibinfo{year}{2006}), \eprint{astro-ph/0603145}.

\bibitem[{\citenamefont{{Anderson} et~al.}(2008)\citenamefont{{Anderson},
  {Hirschmann}, {Lehner}, {Liebling}, {Motl}, {Neilsen}, {Palenzuela}, and
  {Tohline}}}]{2008PhRvL.100s1101A}
\bibinfo{author}{\bibfnamefont{M.}~\bibnamefont{{Anderson}}},
  \bibinfo{author}{\bibfnamefont{E.~W.} \bibnamefont{{Hirschmann}}},
  \bibinfo{author}{\bibfnamefont{L.}~\bibnamefont{{Lehner}}},
  \bibinfo{author}{\bibfnamefont{S.~L.} \bibnamefont{{Liebling}}},
  \bibinfo{author}{\bibfnamefont{P.~M.} \bibnamefont{{Motl}}},
  \bibinfo{author}{\bibfnamefont{D.}~\bibnamefont{{Neilsen}}},
  \bibinfo{author}{\bibfnamefont{C.}~\bibnamefont{{Palenzuela}}},
  \bibnamefont{and} \bibinfo{author}{\bibfnamefont{J.~E.}
  \bibnamefont{{Tohline}}}, \bibinfo{journal}{Physical Review Letters}
  \textbf{\bibinfo{volume}{100}}, \bibinfo{pages}{191101}
  (\bibinfo{year}{2008}), \eprint{gr-qc/0801.4387}.

\bibitem[{\citenamefont{{Liu} et~al.}(2008)\citenamefont{{Liu}, {Shapiro},
  {Etienne}, and {Taniguchi}}}]{2008PhRvD..78b4012L}
\bibinfo{author}{\bibfnamefont{Y.~T.} \bibnamefont{{Liu}}},
  \bibinfo{author}{\bibfnamefont{S.~L.} \bibnamefont{{Shapiro}}},
  \bibinfo{author}{\bibfnamefont{Z.~B.} \bibnamefont{{Etienne}}},
  \bibnamefont{and}
  \bibinfo{author}{\bibfnamefont{K.}~\bibnamefont{{Taniguchi}}},
  \bibinfo{journal}{\prd} \textbf{\bibinfo{volume}{78}},
  \bibinfo{pages}{024012} (\bibinfo{year}{2008}), \eprint{gr-qc/0803.4193}.

\bibitem[{\citenamefont{{Lee} and
  {Klu{\'z}niak}}(1999{\natexlab{a}})}]{1999MNRAS.308..780L}
\bibinfo{author}{\bibfnamefont{W.~H.} \bibnamefont{{Lee}}} \bibnamefont{and}
  \bibinfo{author}{\bibfnamefont{W.~{\L}.} \bibnamefont{{Klu{\'z}niak}}},
  \bibinfo{journal}{MNRAS} \textbf{\bibinfo{volume}{308}}, \bibinfo{pages}{780}
  (\bibinfo{year}{1999}{\natexlab{a}}), \eprint{astro-ph/9904328}.

\bibitem[{\citenamefont{{Lee} and
  {Klu{\'z}niak}}(1999{\natexlab{b}})}]{1999ApJ...526..178L}
\bibinfo{author}{\bibfnamefont{W.~H.} \bibnamefont{{Lee}}} \bibnamefont{and}
  \bibinfo{author}{\bibfnamefont{W.~{\L}.} \bibnamefont{{Klu{\'z}niak}}},
  \bibinfo{journal}{Astrophys.\ J.} \textbf{\bibinfo{volume}{526}},
  \bibinfo{pages}{178} (\bibinfo{year}{1999}{\natexlab{b}}),
  \eprint{astro-ph/9808185}.

\bibitem[{\citenamefont{{Janka} et~al.}(1999)\citenamefont{{Janka}, {Eberl},
  {Ruffert}, and {Fryer}}}]{1999ApJ...527L..39J}
\bibinfo{author}{\bibfnamefont{H.-T.} \bibnamefont{{Janka}}},
  \bibinfo{author}{\bibfnamefont{T.}~\bibnamefont{{Eberl}}},
  \bibinfo{author}{\bibfnamefont{M.}~\bibnamefont{{Ruffert}}},
  \bibnamefont{and} \bibinfo{author}{\bibfnamefont{C.~L.}
  \bibnamefont{{Fryer}}}, \bibinfo{journal}{Astrophys.\ J.\ Letters}
  \textbf{\bibinfo{volume}{527}}, \bibinfo{pages}{L39} (\bibinfo{year}{1999}),
  \eprint{astro-ph/9908290}.

\bibitem[{\citenamefont{{Rosswog} et~al.}(2004)\citenamefont{{Rosswog},
  {Speith}, and {Wynn}}}]{2004MNRAS.351.1121R}
\bibinfo{author}{\bibfnamefont{S.}~\bibnamefont{{Rosswog}}},
  \bibinfo{author}{\bibfnamefont{R.}~\bibnamefont{{Speith}}}, \bibnamefont{and}
  \bibinfo{author}{\bibfnamefont{G.~A.} \bibnamefont{{Wynn}}},
  \bibinfo{journal}{MNRAS} \textbf{\bibinfo{volume}{351}},
  \bibinfo{pages}{1121} (\bibinfo{year}{2004}), \eprint{astro-ph/0403500}.

\bibitem[{\citenamefont{{Faber} et~al.}(2006)\citenamefont{{Faber},
  {Baumgarte}, {Shapiro}, {Taniguchi}, and {Rasio}}}]{2006PhRvD..73b4012F}
\bibinfo{author}{\bibfnamefont{J.~A.} \bibnamefont{{Faber}}},
  \bibinfo{author}{\bibfnamefont{T.~W.} \bibnamefont{{Baumgarte}}},
  \bibinfo{author}{\bibfnamefont{S.~L.} \bibnamefont{{Shapiro}}},
  \bibinfo{author}{\bibfnamefont{K.}~\bibnamefont{{Taniguchi}}},
  \bibnamefont{and} \bibinfo{author}{\bibfnamefont{F.~A.}
  \bibnamefont{{Rasio}}}, \bibinfo{journal}{Phys.\ Rev.\ D}
  \textbf{\bibinfo{volume}{73}}, \bibinfo{pages}{024012}
  (\bibinfo{year}{2006}), \eprint{astro-ph/0511366}.

\bibitem[{\citenamefont{{L{\"o}ffler} et~al.}(2006)\citenamefont{{L{\"o}ffler},
  {Rezzolla}, and {Ansorg}}}]{Loffler:2006wa}
\bibinfo{author}{\bibfnamefont{F.}~\bibnamefont{{L{\"o}ffler}}},
  \bibinfo{author}{\bibfnamefont{L.}~\bibnamefont{{Rezzolla}}},
  \bibnamefont{and} \bibinfo{author}{\bibfnamefont{M.}~\bibnamefont{{Ansorg}}}
  (\bibinfo{year}{2006}), \eprint{gr-qc/0606104}.

\bibitem[{\citenamefont{{Shibata} and {Ury{\= u}}}(2007)}]{2007CQGra..24..125S}
\bibinfo{author}{\bibfnamefont{M.}~\bibnamefont{{Shibata}}} \bibnamefont{and}
  \bibinfo{author}{\bibfnamefont{K.}~\bibnamefont{{Ury{\= u}}}},
  \bibinfo{journal}{Classical and Quantum Gravity}
  \textbf{\bibinfo{volume}{24}}, \bibinfo{pages}{125} (\bibinfo{year}{2007}),
  \eprint{astro-ph/0611522}.

\bibitem[{\citenamefont{Shibata and Taniguchi}(2007)}]{Shibata:2007zm}
\bibinfo{author}{\bibfnamefont{M.}~\bibnamefont{Shibata}} \bibnamefont{and}
  \bibinfo{author}{\bibfnamefont{K.}~\bibnamefont{Taniguchi}}
  (\bibinfo{year}{2007}), \eprint{gr-qc/0711.1410}.

\bibitem[{\citenamefont{Yamamoto et~al.}(2008)\citenamefont{Yamamoto, Shibata,
  and Taniguchi}}]{Yamamoto:2008js}
\bibinfo{author}{\bibfnamefont{T.}~\bibnamefont{Yamamoto}},
  \bibinfo{author}{\bibfnamefont{M.}~\bibnamefont{Shibata}}, \bibnamefont{and}
  \bibinfo{author}{\bibfnamefont{K.}~\bibnamefont{Taniguchi}}
  (\bibinfo{year}{2008}), \eprint{gr-qc/0806.4007}.

\bibitem[{\citenamefont{{Etienne} et~al.}(2008)\citenamefont{{Etienne},
  {Faber}, {Liu}, {Shapiro}, {Taniguchi}, and
  {Baumgarte}}}]{2008PhRvD..77h4002E}
\bibinfo{author}{\bibfnamefont{Z.~B.} \bibnamefont{{Etienne}}},
  \bibinfo{author}{\bibfnamefont{J.~A.} \bibnamefont{{Faber}}},
  \bibinfo{author}{\bibfnamefont{Y.~T.} \bibnamefont{{Liu}}},
  \bibinfo{author}{\bibfnamefont{S.~L.} \bibnamefont{{Shapiro}}},
  \bibinfo{author}{\bibfnamefont{K.}~\bibnamefont{{Taniguchi}}},
  \bibnamefont{and} \bibinfo{author}{\bibfnamefont{T.~W.}
  \bibnamefont{{Baumgarte}}}, \bibinfo{journal}{\prd}
  \textbf{\bibinfo{volume}{77}}, \bibinfo{pages}{084002}
  (\bibinfo{year}{2008}), \eprint{astro-ph/0712.2460}.

\bibitem[{\citenamefont{Kidder et~al.}(2000)\citenamefont{Kidder, Scheel,
  Teukolsky, Carlson, and Cook}}]{Kidder:2000yq}
\bibinfo{author}{\bibfnamefont{L.~E.} \bibnamefont{Kidder}},
  \bibinfo{author}{\bibfnamefont{M.~A.} \bibnamefont{Scheel}},
  \bibinfo{author}{\bibfnamefont{S.~A.} \bibnamefont{Teukolsky}},
  \bibinfo{author}{\bibfnamefont{E.~D.} \bibnamefont{Carlson}},
  \bibnamefont{and} \bibinfo{author}{\bibfnamefont{G.~B.} \bibnamefont{Cook}},
  \bibinfo{journal}{Phys. Rev.} \textbf{\bibinfo{volume}{D62}},
  \bibinfo{pages}{084032} (\bibinfo{year}{2000}), \eprint{gr-qc/0005056}.

\bibitem[{\citenamefont{Scheel et~al.}(2002)\citenamefont{Scheel, Kidder,
  Lindblom, Pfeiffer, and Teukolsky}}]{Scheel:2002yj}
\bibinfo{author}{\bibfnamefont{M.~A.} \bibnamefont{Scheel}},
  \bibinfo{author}{\bibfnamefont{L.~E.} \bibnamefont{Kidder}},
  \bibinfo{author}{\bibfnamefont{L.}~\bibnamefont{Lindblom}},
  \bibinfo{author}{\bibfnamefont{H.~P.} \bibnamefont{Pfeiffer}},
  \bibnamefont{and} \bibinfo{author}{\bibfnamefont{S.~A.}
  \bibnamefont{Teukolsky}}, \bibinfo{journal}{Phys. Rev.}
  \textbf{\bibinfo{volume}{D66}}, \bibinfo{pages}{124005}
  (\bibinfo{year}{2002}), \eprint{gr-qc/0209115}.

\bibitem[{\citenamefont{Lindblom et~al.}(2006)\citenamefont{Lindblom, Scheel,
  Kidder, Owen, and Rinne}}]{Lindblom:2005qh}
\bibinfo{author}{\bibfnamefont{L.}~\bibnamefont{Lindblom}},
  \bibinfo{author}{\bibfnamefont{M.~A.} \bibnamefont{Scheel}},
  \bibinfo{author}{\bibfnamefont{L.~E.} \bibnamefont{Kidder}},
  \bibinfo{author}{\bibfnamefont{R.}~\bibnamefont{Owen}}, \bibnamefont{and}
  \bibinfo{author}{\bibfnamefont{O.}~\bibnamefont{Rinne}},
  \bibinfo{journal}{Class. Quant. Grav.} \textbf{\bibinfo{volume}{23}},
  \bibinfo{pages}{S447} (\bibinfo{year}{2006}), \eprint{gr-qc/0512093}.

\bibitem[{\citenamefont{Boyle et~al.}(2007{\natexlab{a}})\citenamefont{Boyle,
  Lindblom, Pfeiffer, Scheel, and Kidder}}]{Boyle:2006ne}
\bibinfo{author}{\bibfnamefont{M.}~\bibnamefont{Boyle}},
  \bibinfo{author}{\bibfnamefont{L.}~\bibnamefont{Lindblom}},
  \bibinfo{author}{\bibfnamefont{H.}~\bibnamefont{Pfeiffer}},
  \bibinfo{author}{\bibfnamefont{M.}~\bibnamefont{Scheel}}, \bibnamefont{and}
  \bibinfo{author}{\bibfnamefont{L.~E.} \bibnamefont{Kidder}},
  \bibinfo{journal}{Phys. Rev.} \textbf{\bibinfo{volume}{D75}},
  \bibinfo{pages}{024006} (\bibinfo{year}{2007}{\natexlab{a}}),
  \eprint{gr-qc/0609047}.

\bibitem[{\citenamefont{Scheel et~al.}(2006)}]{Scheel:2006gg}
\bibinfo{author}{\bibfnamefont{M.~A.} \bibnamefont{Scheel}}
  \bibnamefont{et~al.}, \bibinfo{journal}{Phys. Rev.}
  \textbf{\bibinfo{volume}{D74}}, \bibinfo{pages}{104006}
  (\bibinfo{year}{2006}), \eprint{gr-qc/0607056}.

\bibitem[{\citenamefont{Pfeiffer et~al.}(2007)}]{Pfeiffer:2007yz}
\bibinfo{author}{\bibfnamefont{H.~P.} \bibnamefont{Pfeiffer}}
  \bibnamefont{et~al.}, \bibinfo{journal}{Class. Quant. Grav.}
  \textbf{\bibinfo{volume}{24}}, \bibinfo{pages}{S59} (\bibinfo{year}{2007}),
  \eprint{gr-qc/0702106}.

\bibitem[{\citenamefont{Boyle et~al.}(2007{\natexlab{b}})}]{Boyle:2007ft}
\bibinfo{author}{\bibfnamefont{M.}~\bibnamefont{Boyle}} \bibnamefont{et~al.},
  \bibinfo{journal}{Phys. Rev.} \textbf{\bibinfo{volume}{D76}},
  \bibinfo{pages}{124038} (\bibinfo{year}{2007}{\natexlab{b}}),
  \eprint{gr-qc/0710.0158}.

\bibitem[{\citenamefont{Tadmor}(1989)}]{t89}
\bibinfo{author}{\bibfnamefont{E.}~\bibnamefont{Tadmor}},
  \bibinfo{journal}{SIAM J.\ Numer.\ Anal.} \textbf{\bibinfo{volume}{26}},
  \bibinfo{pages}{30} (\bibinfo{year}{1989}).

\bibitem[{\citenamefont{Don}(1994)}]{d94}
\bibinfo{author}{\bibfnamefont{W.~S.} \bibnamefont{Don}}, \bibinfo{journal}{J.\
  Comput.\ Phys.} \textbf{\bibinfo{volume}{110}}, \bibinfo{pages}{103}
  (\bibinfo{year}{1994}).

\bibitem[{\citenamefont{{Dimmelmeier} et~al.}(2005)\citenamefont{{Dimmelmeier},
  {Novak}, {Font}, {Ib{\'a}{\~n}ez}, and {M{\"u}ller}}}]{2005PhRvD..71f4023D}
\bibinfo{author}{\bibfnamefont{H.}~\bibnamefont{{Dimmelmeier}}},
  \bibinfo{author}{\bibfnamefont{J.}~\bibnamefont{{Novak}}},
  \bibinfo{author}{\bibfnamefont{J.~A.} \bibnamefont{{Font}}},
  \bibinfo{author}{\bibfnamefont{J.~M.} \bibnamefont{{Ib{\'a}{\~n}ez}}},
  \bibnamefont{and}
  \bibinfo{author}{\bibfnamefont{E.}~\bibnamefont{{M{\"u}ller}}},
  \bibinfo{journal}{Phys.\ Rev.\ D} \textbf{\bibinfo{volume}{71}},
  \bibinfo{pages}{064023} (\bibinfo{year}{2005}), \eprint{astro-ph/0407174}.

\bibitem[{\citenamefont{Dimmelmeier et~al.}(2006)\citenamefont{Dimmelmeier,
  Cerda-Duran, and Marek}}]{Dimmelmeier:2006mh}
\bibinfo{author}{\bibfnamefont{H.}~\bibnamefont{Dimmelmeier}},
  \bibinfo{author}{\bibfnamefont{P.}~\bibnamefont{Cerda-Duran}},
  \bibnamefont{and} \bibinfo{author}{\bibfnamefont{A.}~\bibnamefont{Marek}},
  \bibinfo{journal}{AIP Conf. Proc.} \textbf{\bibinfo{volume}{861}},
  \bibinfo{pages}{596} (\bibinfo{year}{2006}), \eprint{astro-ph/0603760}.

\bibitem[{\citenamefont{Ott et~al.}(2007)}]{Ott:2006eu}
\bibinfo{author}{\bibfnamefont{C.~D.} \bibnamefont{Ott}} \bibnamefont{et~al.},
  \bibinfo{journal}{Phys. Rev. Lett.} \textbf{\bibinfo{volume}{98}},
  \bibinfo{pages}{261101} (\bibinfo{year}{2007}), \eprint{astro-ph/0609819}.

\bibitem[{\citenamefont{Dimmelmeier et~al.}(2007)\citenamefont{Dimmelmeier,
  Ott, Janka, Marek, and Mueller}}]{Dimmelmeier:2007ui}
\bibinfo{author}{\bibfnamefont{H.}~\bibnamefont{Dimmelmeier}},
  \bibinfo{author}{\bibfnamefont{C.~D.} \bibnamefont{Ott}},
  \bibinfo{author}{\bibfnamefont{H.-T.} \bibnamefont{Janka}},
  \bibinfo{author}{\bibfnamefont{A.}~\bibnamefont{Marek}}, \bibnamefont{and}
  \bibinfo{author}{\bibfnamefont{E.}~\bibnamefont{Mueller}},
  \bibinfo{journal}{Phys. Rev. Lett.} \textbf{\bibinfo{volume}{98}},
  \bibinfo{pages}{251101} (\bibinfo{year}{2007}), \eprint{astro-ph/0702305}.

\bibitem[{\citenamefont{Cerda-Duran et~al.}(2007)\citenamefont{Cerda-Duran,
  Font, and Dimmelmeier}}]{CerdaDuran:2007cr}
\bibinfo{author}{\bibfnamefont{P.}~\bibnamefont{Cerda-Duran}},
  \bibinfo{author}{\bibfnamefont{J.~A.} \bibnamefont{Font}}, \bibnamefont{and}
  \bibinfo{author}{\bibfnamefont{H.}~\bibnamefont{Dimmelmeier}}
  (\bibinfo{year}{2007}), \eprint{astro-ph/0703597}.

\bibitem[{\citenamefont{{Friedrich}}(1985)}]{1985CMaPh.100..525F}
\bibinfo{author}{\bibfnamefont{H.}~\bibnamefont{{Friedrich}}},
  \bibinfo{journal}{Communications in Mathematical Physics}
  \textbf{\bibinfo{volume}{100}}, \bibinfo{pages}{525} (\bibinfo{year}{1985}).

\bibitem[{\citenamefont{Pretorius}(2005)}]{Pretorius:2004jg}
\bibinfo{author}{\bibfnamefont{F.}~\bibnamefont{Pretorius}},
  \bibinfo{journal}{Class. Quant. Grav.} \textbf{\bibinfo{volume}{22}},
  \bibinfo{pages}{425} (\bibinfo{year}{2005}), \eprint{gr-qc/0407110}.

\bibitem[{\citenamefont{Lehner et~al.}(2005)\citenamefont{Lehner, Reula, and
  Tiglio}}]{Lehner:2005bz}
\bibinfo{author}{\bibfnamefont{L.}~\bibnamefont{Lehner}},
  \bibinfo{author}{\bibfnamefont{O.}~\bibnamefont{Reula}}, \bibnamefont{and}
  \bibinfo{author}{\bibfnamefont{M.}~\bibnamefont{Tiglio}},
  \bibinfo{journal}{Class. Quant. Grav.} \textbf{\bibinfo{volume}{22}},
  \bibinfo{pages}{5283} (\bibinfo{year}{2005}), \eprint{gr-qc/0507004}.

\bibitem[{\citenamefont{Matsushima and Marcus}(1995)}]{mm95}
\bibinfo{author}{\bibfnamefont{T.}~\bibnamefont{Matsushima}} \bibnamefont{and}
  \bibinfo{author}{\bibfnamefont{P.~S.} \bibnamefont{Marcus}},
  \bibinfo{journal}{Journal of Computational Physics}
  \textbf{\bibinfo{volume}{120}}, \bibinfo{pages}{365} (\bibinfo{year}{1995}).

\bibitem[{\citenamefont{Rinne et~al.}(2007)\citenamefont{Rinne, Lindblom, and
  Scheel}}]{Rinne:2007ui}
\bibinfo{author}{\bibfnamefont{O.}~\bibnamefont{Rinne}},
  \bibinfo{author}{\bibfnamefont{L.}~\bibnamefont{Lindblom}}, \bibnamefont{and}
  \bibinfo{author}{\bibfnamefont{M.~A.} \bibnamefont{Scheel}},
  \bibinfo{journal}{Class. Quant. Grav.} \textbf{\bibinfo{volume}{24}},
  \bibinfo{pages}{4053} (\bibinfo{year}{2007}), \eprint{gr-qc/0704.0782}.

\bibitem[{\citenamefont{Boyd}(1992)}]{b92}
\bibinfo{author}{\bibfnamefont{J.~P.} \bibnamefont{Boyd}},
  \bibinfo{journal}{Journal of Computational Physics}
  \textbf{\bibinfo{volume}{103}}, \bibinfo{pages}{243} (\bibinfo{year}{1992}).

\bibitem[{\citenamefont{{Font} et~al.}(2002)\citenamefont{{Font}, {Goodale},
  {Iyer}, {Miller}, {Rezzolla}, {Seidel}, {Stergioulas}, {Suen}, and
  {Tobias}}}]{2002PhRvD..65h4024F}
\bibinfo{author}{\bibfnamefont{J.~A.} \bibnamefont{{Font}}},
  \bibinfo{author}{\bibfnamefont{T.}~\bibnamefont{{Goodale}}},
  \bibinfo{author}{\bibfnamefont{S.}~\bibnamefont{{Iyer}}},
  \bibinfo{author}{\bibfnamefont{M.}~\bibnamefont{{Miller}}},
  \bibinfo{author}{\bibfnamefont{L.}~\bibnamefont{{Rezzolla}}},
  \bibinfo{author}{\bibfnamefont{E.}~\bibnamefont{{Seidel}}},
  \bibinfo{author}{\bibfnamefont{N.}~\bibnamefont{{Stergioulas}}},
  \bibinfo{author}{\bibfnamefont{W.-M.} \bibnamefont{{Suen}}},
  \bibnamefont{and} \bibinfo{author}{\bibfnamefont{M.}~\bibnamefont{{Tobias}}},
  \bibinfo{journal}{\prd} \textbf{\bibinfo{volume}{65}},
  \bibinfo{pages}{084024} (\bibinfo{year}{2002}), \eprint{gr-qc/0110047}.

\bibitem[{\citenamefont{{Baiotti} et~al.}(2005)\citenamefont{{Baiotti},
  {Hawke}, {Montero}, {L{\"o}ffler}, {Rezzolla}, {Stergioulas}, {Font}, and
  {Seidel}}}]{2005PhRvD..71b4035B}
\bibinfo{author}{\bibfnamefont{L.}~\bibnamefont{{Baiotti}}},
  \bibinfo{author}{\bibfnamefont{I.}~\bibnamefont{{Hawke}}},
  \bibinfo{author}{\bibfnamefont{P.~J.} \bibnamefont{{Montero}}},
  \bibinfo{author}{\bibfnamefont{F.}~\bibnamefont{{L{\"o}ffler}}},
  \bibinfo{author}{\bibfnamefont{L.}~\bibnamefont{{Rezzolla}}},
  \bibinfo{author}{\bibfnamefont{N.}~\bibnamefont{{Stergioulas}}},
  \bibinfo{author}{\bibfnamefont{J.~A.} \bibnamefont{{Font}}},
  \bibnamefont{and} \bibinfo{author}{\bibfnamefont{E.}~\bibnamefont{{Seidel}}},
  \bibinfo{journal}{\prd} \textbf{\bibinfo{volume}{71}},
  \bibinfo{pages}{024035} (\bibinfo{year}{2005}), \eprint{gr-qc/0403029}.

\bibitem[{\citenamefont{{Colella} and {Woodward}}(1984)}]{1984JCoPh..54..174C}
\bibinfo{author}{\bibfnamefont{P.}~\bibnamefont{{Colella}}} \bibnamefont{and}
  \bibinfo{author}{\bibfnamefont{P.~R.} \bibnamefont{{Woodward}}},
  \bibinfo{journal}{Journal of Computational Physics}
  \textbf{\bibinfo{volume}{54}}, \bibinfo{pages}{174} (\bibinfo{year}{1984}).

\bibitem[{\citenamefont{Liu and Osher}(1998)}]{lo98}
\bibinfo{author}{\bibfnamefont{X.-D.} \bibnamefont{Liu}} \bibnamefont{and}
  \bibinfo{author}{\bibfnamefont{S.}~\bibnamefont{Osher}},
  \bibinfo{journal}{Journal of Computational Physics}
  \textbf{\bibinfo{volume}{142}}, \bibinfo{pages}{304} (\bibinfo{year}{1998}).

\bibitem[{\citenamefont{Xu-Dong~Liu}(1994)}]{loc94}
\bibinfo{author}{\bibfnamefont{T.~C.} \bibnamefont{Xu-Dong~Liu},
  \bibfnamefont{Stanley~Osher}}, \bibinfo{journal}{Journal of Computational
  Physics} \textbf{\bibinfo{volume}{115}}, \bibinfo{pages}{213}
  (\bibinfo{year}{1994}).

\bibitem[{\citenamefont{A.~Harten}(1983)}]{hll}
\bibinfo{author}{\bibfnamefont{B.~v.~L.} \bibnamefont{A.~Harten},
  \bibfnamefont{P.~D.~Lax}}, \bibinfo{journal}{SIAM Rev.}
  \textbf{\bibinfo{volume}{25}}, \bibinfo{pages}{35} (\bibinfo{year}{1983}).

\bibitem[{\citenamefont{{Del Zanna} and
  {Bucciantini}}(2002)}]{2002A&A...390.1177D}
\bibinfo{author}{\bibfnamefont{L.}~\bibnamefont{{Del Zanna}}} \bibnamefont{and}
  \bibinfo{author}{\bibfnamefont{N.}~\bibnamefont{{Bucciantini}}},
  \bibinfo{journal}{Astron.\ and Astrophys.} \textbf{\bibinfo{volume}{390}},
  \bibinfo{pages}{1177} (\bibinfo{year}{2002}), \eprint{astro-ph/0205290}.

\bibitem[{\citenamefont{Hawke et~al.}(2005)\citenamefont{Hawke, Loffler, and
  Nerozzi}}]{Hawke:2005zw}
\bibinfo{author}{\bibfnamefont{I.}~\bibnamefont{Hawke}},
  \bibinfo{author}{\bibfnamefont{F.}~\bibnamefont{Loffler}}, \bibnamefont{and}
  \bibinfo{author}{\bibfnamefont{A.}~\bibnamefont{Nerozzi}},
  \bibinfo{journal}{Phys. Rev.} \textbf{\bibinfo{volume}{D71}},
  \bibinfo{pages}{104006} (\bibinfo{year}{2005}), \eprint{gr-qc/0501054}.

\bibitem[{\citenamefont{{Michel}}(1972)}]{1972Ap&SS..15..153M}
\bibinfo{author}{\bibfnamefont{F.~C.} \bibnamefont{{Michel}}},
  \bibinfo{journal}{Astrophys. Scapse Sci.} \textbf{\bibinfo{volume}{15}},
  \bibinfo{pages}{153} (\bibinfo{year}{1972}).

\bibitem[{\citenamefont{{Petrich} et~al.}(1988)\citenamefont{{Petrich},
  {Shapiro}, and {Teukolsky}}}]{1988PhRvL..60.1781P}
\bibinfo{author}{\bibfnamefont{L.~I.} \bibnamefont{{Petrich}}},
  \bibinfo{author}{\bibfnamefont{S.~L.} \bibnamefont{{Shapiro}}},
  \bibnamefont{and} \bibinfo{author}{\bibfnamefont{S.~A.}
  \bibnamefont{{Teukolsky}}}, \bibinfo{journal}{Physical Review Letters}
  \textbf{\bibinfo{volume}{60}}, \bibinfo{pages}{1781} (\bibinfo{year}{1988}).

\bibitem[{\citenamefont{{Cook} et~al.}(1992)\citenamefont{{Cook}, {Shapiro},
  and {Teukolsky}}}]{1992ApJ...398..203C}
\bibinfo{author}{\bibfnamefont{G.~B.} \bibnamefont{{Cook}}},
  \bibinfo{author}{\bibfnamefont{S.~L.} \bibnamefont{{Shapiro}}},
  \bibnamefont{and} \bibinfo{author}{\bibfnamefont{S.~A.}
  \bibnamefont{{Teukolsky}}}, \bibinfo{journal}{\apj}
  \textbf{\bibinfo{volume}{398}}, \bibinfo{pages}{203} (\bibinfo{year}{1992}).

\bibitem[{\citenamefont{Friedman et~al.}(1988)\citenamefont{Friedman, Ipser,
  and Sorkin}}]{Friedman:1988er}
\bibinfo{author}{\bibfnamefont{J.~L.} \bibnamefont{Friedman}},
  \bibinfo{author}{\bibfnamefont{J.~R.} \bibnamefont{Ipser}}, \bibnamefont{and}
  \bibinfo{author}{\bibfnamefont{R.~D.} \bibnamefont{Sorkin}},
  \bibinfo{journal}{Astrophys. J.} \textbf{\bibinfo{volume}{325}},
  \bibinfo{pages}{722} (\bibinfo{year}{1988}).

\bibitem[{\citenamefont{Baumgarte et~al.}(2000)\citenamefont{Baumgarte,
  Shapiro, and Shibata}}]{Baumgarte:1999cq}
\bibinfo{author}{\bibfnamefont{T.~W.} \bibnamefont{Baumgarte}},
  \bibinfo{author}{\bibfnamefont{S.~L.} \bibnamefont{Shapiro}},
  \bibnamefont{and} \bibinfo{author}{\bibfnamefont{M.}~\bibnamefont{Shibata}},
  \bibinfo{journal}{Astrophys. J.} \textbf{\bibinfo{volume}{528}},
  \bibinfo{pages}{L29} (\bibinfo{year}{2000}), \eprint{astro-ph/9910565}.

\bibitem[{\citenamefont{Duez et~al.}(2005)\citenamefont{Duez, Liu, Shapiro, and
  Stephens}}]{Duez:2005sf}
\bibinfo{author}{\bibfnamefont{M.~D.} \bibnamefont{Duez}},
  \bibinfo{author}{\bibfnamefont{Y.~T.} \bibnamefont{Liu}},
  \bibinfo{author}{\bibfnamefont{S.~L.} \bibnamefont{Shapiro}},
  \bibnamefont{and} \bibinfo{author}{\bibfnamefont{B.~C.}
  \bibnamefont{Stephens}}, \bibinfo{journal}{Phys. Rev.}
  \textbf{\bibinfo{volume}{D72}}, \bibinfo{pages}{024028}
  (\bibinfo{year}{2005}), \eprint{astro-ph/0503420}.

\bibitem[{\citenamefont{{Douchin} and {Haensel}}(2001)}]{2001A&A...380..151D}
\bibinfo{author}{\bibfnamefont{F.}~\bibnamefont{{Douchin}}} \bibnamefont{and}
  \bibinfo{author}{\bibfnamefont{P.}~\bibnamefont{{Haensel}}},
  \bibinfo{journal}{Astron.\ and Astrophys.} \textbf{\bibinfo{volume}{380}},
  \bibinfo{pages}{151} (\bibinfo{year}{2001}), \eprint{astro-ph/0111092}.

\bibitem[{\citenamefont{Foucart et~al.}(2008)\citenamefont{Foucart, Kidder,
  Pfeiffer, and Teukolsky}}]{Foucart:2008qt}
\bibinfo{author}{\bibfnamefont{F.}~\bibnamefont{Foucart}},
  \bibinfo{author}{\bibfnamefont{L.~E.} \bibnamefont{Kidder}},
  \bibinfo{author}{\bibfnamefont{H.~P.} \bibnamefont{Pfeiffer}},
  \bibnamefont{and} \bibinfo{author}{\bibfnamefont{S.~A.}
  \bibnamefont{Teukolsky}}, \bibinfo{journal}{Phys. Rev.}
  \textbf{\bibinfo{volume}{D77}}, \bibinfo{pages}{124051}
  (\bibinfo{year}{2008}), \eprint{gr-qc/0804.3787}.

\bibitem[{\citenamefont{Caudill et~al.}(2006)\citenamefont{Caudill, Cook,
  Grigsby, and Pfeiffer}}]{Caudill:2006hw}
\bibinfo{author}{\bibfnamefont{M.}~\bibnamefont{Caudill}},
  \bibinfo{author}{\bibfnamefont{G.~B.} \bibnamefont{Cook}},
  \bibinfo{author}{\bibfnamefont{J.~D.} \bibnamefont{Grigsby}},
  \bibnamefont{and} \bibinfo{author}{\bibfnamefont{H.~P.}
  \bibnamefont{Pfeiffer}}, \bibinfo{journal}{Phys. Rev.}
  \textbf{\bibinfo{volume}{D74}}, \bibinfo{pages}{064011}
  (\bibinfo{year}{2006}), \eprint{gr-qc/0605053}.

\bibitem[{\citenamefont{Lovelace et~al.}(2008)\citenamefont{Lovelace, Owen,
  Pfeiffer, and Chu}}]{Lovelace:2008tw}
\bibinfo{author}{\bibfnamefont{G.}~\bibnamefont{Lovelace}},
  \bibinfo{author}{\bibfnamefont{R.}~\bibnamefont{Owen}},
  \bibinfo{author}{\bibfnamefont{H.~P.} \bibnamefont{Pfeiffer}},
  \bibnamefont{and} \bibinfo{author}{\bibfnamefont{T.}~\bibnamefont{Chu}}
  (\bibinfo{year}{2008}), \eprint{gr-qc/0805.4192}.

\bibitem[{\citenamefont{Catlett et~al.}(2007)}]{teragrid}
\bibinfo{author}{\bibfnamefont{C.}~\bibnamefont{Catlett}} \bibnamefont{et~al.},
  in \emph{\bibinfo{booktitle}{Advances in Parallel Computing}}, edited by
  \bibinfo{editor}{\bibfnamefont{L.}~\bibnamefont{Grandinetti}}
  (\bibinfo{publisher}{IOS press, Amsterdam}, \bibinfo{year}{2007}).

\end{thebibliography}

\end{document}